\documentclass[aps,twocolumn,showpacs,preprintnumbers,superscriptaddress]{revtex4}
\usepackage{amsmath}
\usepackage{amssymb}
\usepackage{mathrsfs}
\usepackage{graphicx}
\usepackage{longtable}
\usepackage{tabularx}
\usepackage{float}
\usepackage{bm}
\usepackage{natbib}
\usepackage{epstopdf}
\usepackage[colorlinks=true,citecolor=blue,linkcolor=blue,urlcolor=blue]{hyperref}

\begin{document}


\title{Tuning the electronic properties of gated multilayer phosphorene: A self-consistent tight-binding study}

\author{L. L. Li}
\email{longlong.li@uantwerpen.be}
\affiliation{Department of Physics, University of Antwerp,
Groenenborgerlaan 171, B-2020 Antwerpen, Belgium}
\author{B. Partoens}
\email{bart.partoens@uantwerpen.be}
\affiliation{Department of Physics, University of Antwerp,
Groenenborgerlaan 171, B-2020 Antwerpen, Belgium}
\author{F. M. Peeters}
\email{francois.peeters@uantwerpen.be}
\affiliation{Department of Physics, University of Antwerp,
Groenenborgerlaan 171, B-2020 Antwerpen, Belgium}

\date{\today}

\begin{abstract}
By taking account of the electric-field-induced charge screening, a self-consistent calculation within the framework of the tight-binding approach is employed to obtain the electronic band structure of gated multilayer phosphorene and the charge densities on the different phosphorene layers. We find charge density and screening anomalies in single-gated multilayer phosphorene and electron-hole bilayers in dual-gated multilayer phosphorene. Due to the unique puckered lattice structure, both intralayer and interlayer charge screenings are important in gated multilayer phosphorene. We find that the electric-field tuning of the band structure of multilayer phosphorene is distinctively different in the presence and absence of charge screening. For instance, it is shown that the unscreened band gap of multilayer phosphorene decreases dramatically with increasing electric-field strength. However, in the presence of charge screening, the magnitude of this band-gap decrease is significantly reduced and the reduction depends strongly on the number of phosphorene layers. Our theoretical results of the band-gap tuning are compared with recent experiments and good agreement is found.
\end{abstract}


\maketitle

\section{Introduction}

Phosphorene is a single layer of black phosphorus (BP), a relatively new two-dimensional (2D) material which was realized experimentally in 2014 \cite{LiBlackphosphorusfieldeffect2014,LiuPhosphoreneUnexplored2D2014,
LuPlasmaassistedfabricationmonolayer2014,
DasTunableTransportGap2014}. Due to its unique properties, this 2D material has drawn a lot of attention from the research community. For instance, it has a reasonably large band gap combined with a relatively high carrier mobility \cite{QiaoHighmobilitytransportanisotropy2014a}, which is very promising for practical electronic applications in e.g. field-effect transistors. Moreover, it has a puckered honeycomb lattice formed due to $sp^3$ hybridization \cite{RodinStrainInducedGapModification2014a}, which gives rise to highly anisotropic electronic, optical and transport properties \cite{QiaoHighmobilitytransportanisotropy2014a}, such as anisotropic effective mass, optical spectrum and electrical mobility. The highly anisotropic optical properties of phosphorene make it very promising for practical optoelectronic applications in e.g. polarization-sensitive photodetectors \cite{YuanPolarizationsensitivebroadbandphotodetector2015a,
YoungbloodWaveguideintegratedblackphosphorus2015a}.

Due to the successful isolation of phosphorene layers from bulk BP, the electronic, optical and transport properties of multilayer phosphorene have been extensively investigated \cite{TranLayercontrolledbandgap2014,
CastellanosGomezIsolationcharacterizationfewlayer2014,
RudenkoQuasiparticlebandstructure2014,
LeiStackingFaultEnriching2016,CakirSignificanteffectstacking2015,
WangRoleInterlayerCoupling2015,DaiBilayerPhosphoreneEffect2014,
Rudenkorealisticdescriptionmultilayer2015}. This is primarily because these properties can be significantly tuned by the number and the type of stacking layers \cite{TranLayercontrolledbandgap2014,CakirSignificanteffectstacking2015}. For instance, the band gap of phosphorene decreases when increasing the number of layers due to the interlayer electronic coupling \cite{TranLayercontrolledbandgap2014}. The application of external strain, electric field, and magnetic field has also a significant influence on the electronic, optical and transport properties of multilayer phosphorene \cite{FeiStrainEngineeringAnisotropicElectrical2014,
LiuSwitchingNormalInsulator2015,
YuanQuantumHalleffect2016,PereiraLandaulevelssinglelayer2015,
WuFieldinduceddiversequantizations2017}. In particular, it was shown that by applying a perpendicular electric field, a semiconductor-to-semimetal transition can be induced in bilayer phosphorene \cite{LiuSwitchingNormalInsulator2015,YuanQuantumHalleffect2016,
PereiraLandaulevelssinglelayer2015, WuFieldinduceddiversequantizations2017}, leading to the appearance of unconventional Dirac fermions with linear energy spectrum and zero-energy Landau levels \cite{YuanQuantumHalleffect2016,WuFieldinduceddiversequantizations2017}.

It is known that applying an external electric field perpendicular to a multilayer system induces a charge redistribution over the stacked layers, which produces an internal electric field that counteracts the externally applied one (i.e., the electric-field-induced charge screening). Although the influence of a perpendicular electric field on the electronic properties of multilayer phosphorene was widely investigated, the electric-field-induced charge screening effect, which was shown to be of significant importance in multilayer graphene \cite{McCannAsymmetrygapelectronic2006,AvetisyanElectricfieldtuning2009,
AvetisyanElectricfieldcontrolband2009,KoshinoGateinducedinterlayerasymmetry2009,
ZhangBandstructureABC2010}, remains poorly understood in multilayer phosphorene. Up to date, there are few studies exploring the screening effect on the electronic properties of multilayer pshophorene in the presence of a perpendicular electric field \cite{JhunElectronicstructurecharged2017, DengEfficientelectricalcontrol2017}. In Ref. \cite{JhunElectronicstructurecharged2017}, the electronic structure of bilayer and trilayer phosphorene was obtained by first-principles calculations, where the screening effect was induced by the additional doping of charges; whereas in Ref. \cite{DengEfficientelectricalcontrol2017}, the band gap of multilayer phosphorene was obtained by tight-binding calculations, where the screening effect was introduced by assuming a band-gap-dependent dielectric constant. However, the charge screening induced by the perpendicular electric field was not included in these studies.

In the present work, we investigate theoretically the electronic properties of gated multilayer phosphorene by taking into account the electric-field-induced charge screening within an experimental set-up, where an electrostatic gating (with top and/or bottom gates) is applied to produce a perpendicular electric field. Our theoretical study is based on the tight-binding (TB) approach to calculate the band structure of gated multilayer phosphorene. A self-consistent Hartree approximation within this TB framework is employed to obtain the gate-induced charge densities on the different layers of phosphorene. Due to the unique puckered lattice structure, we found both intralayer and interlayer charge screenings that are present in gated multilayer phosphorene, which is different from the result observed in gated multilayer graphene, where only interlayer charge screening is present \cite{McCannAsymmetrygapelectronic2006,AvetisyanElectricfieldtuning2009,
AvetisyanElectricfieldcontrolband2009,KoshinoGateinducedinterlayerasymmetry2009,
ZhangBandstructureABC2010}. We also found that the electric-field tuning of the band structure of multilayer phosphorene is distinctively different in the absence and presence of charge screening. For instance, it is shown that the unscreened band gap of multilayer phosphorene decreases dramatically with increasing field strength. However, in the presence of charge screening, the magnitude of this band-gap decrease is significantly reduced and the reduction is more significant for the case of larger number of phosphorene layers. Moreover, we observe electron-hole bilayers in symmetrically dual-gated multilayer phosphorene with tunable layer-dependent electron/hole densities. This could be interesting for the exploration of electrically tunable Wigner crystallization, charge density waves, excitonic condensation and superfluidity in multilayer phosphorene \cite{ZareniaInhomogeneousphasescoupled2017}.

This paper is organized as follows. In Sec. II, we present the self-consistent TB approach for multilayer phosphorene in the presence of an electrostatic gating. In Sec. III, the main results are presented and analyzed for the electronic properties of gated multilayer phosphorene. We also present a comparison with recent experiments on the electric-field tuning of the multilayer band gap. Finally, we make a summary and give concluding remarks in Sec. IV.

\begin{figure}[htb!]
\begin{center}
\includegraphics[width=0.45\textwidth]{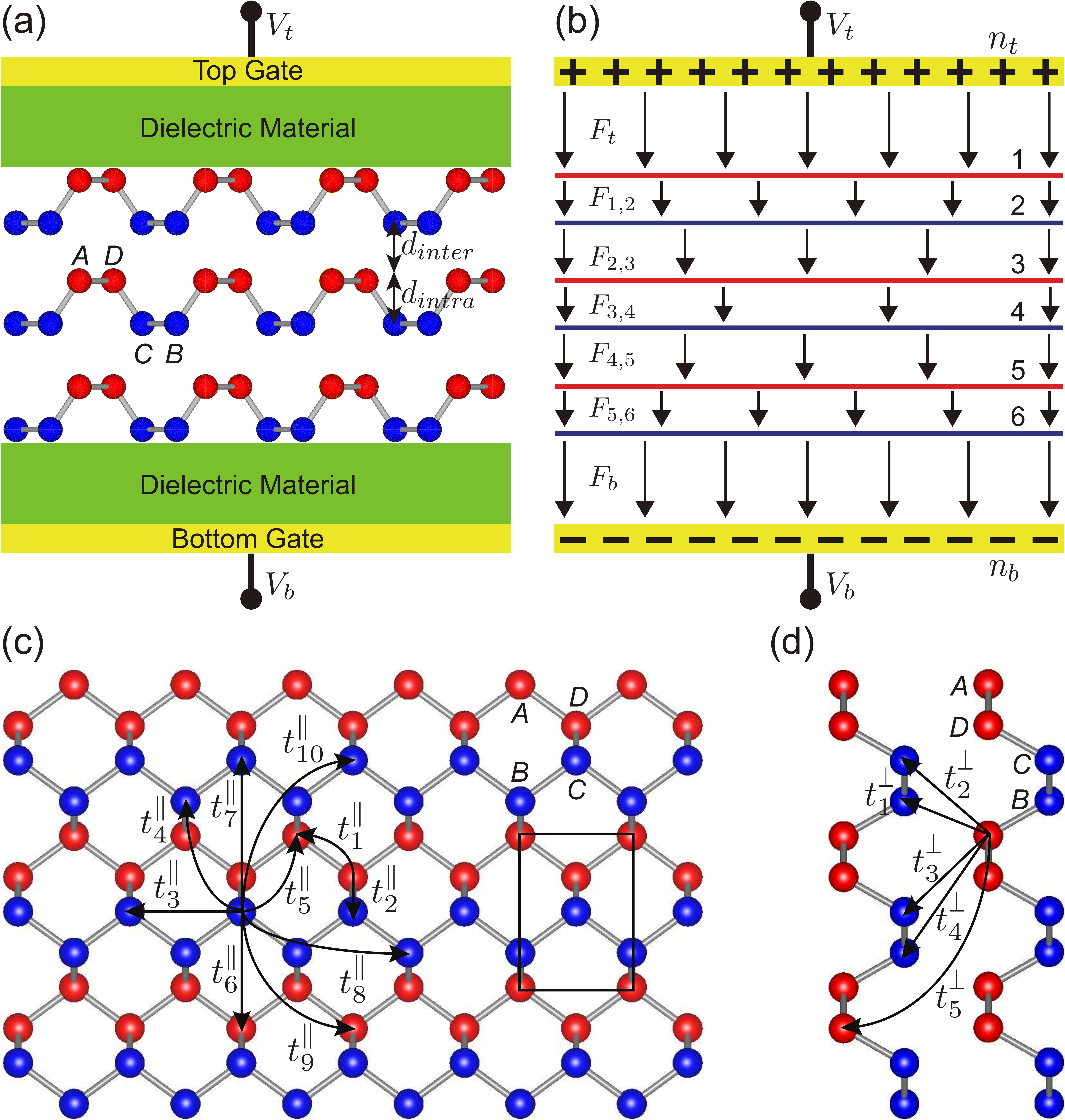}
\caption{(a) and (b) Sketch of the trilayer phosphorene system, on the top (bottom) of which a positively (negatively) charged gate $V_t$ ($V_b$) with charge density $n_t>0$ ($n_b<0$) is placed. (c) and (d) Illustration of the ten intralayer hopping parameters $t_{i}^{\|}$ ($i=1, 2, ..., 10$) and the five interlayer hopping parameters $t_{i}^{\perp}$ ($i=1, 2, ..., 5$) used in the TB model, where the rectangle indicates the unit cell of phosphorene. Due to the puckered lattice structure, trilayer phosphorene has six atomic sublayers depicted by red ($A$, $D$) and blue ($B$, $C$) atoms, where the distance between the two adjacent sublayers is equal to either the interlayer separation ($d_{inter}$) or the intralayer one ($d_{intra}$). The top and bottom gates induce a total carrier density distributed over the different sublayers, i.e., $n=n_t+n_b=\sum_{i=1}^6n_i$, with $n_i$ the carrier density on the $i$-th sublayer. $F_t$ ($F_b$) is the electric field produced by the top (bottom) gate, and $F_{i, i+1}$ ($i=1, 2, ..., 5$) is the total electric field between the two adjacent sublayers.}\label{fig1}
\end{center}
\end{figure}

\section{Self-consistent TB model}

Multilayer phosphorene is modeled as $N$ coupled phosphorene layers which are $AB$ stacked on top of each other, as shown in Fig. \ref{fig1}. From density-functional-theory (DFT) calculations \cite{CakirSignificanteffectstacking2015}, we know that this type of layer stacking is energetically the most stable for bilayer and trilayer phosphorene. In the presence of an electrostatic gating, low-energy electrons and holes in this $N$-layer system are described by the following TB Hamiltonian
\begin{equation}\label{e1}
H=\sum_{i}\varepsilon_ic_i^{\dag}c_i+\sum_{i\neq j}t_{ij}^{\|}c_i^{\dag}c_j+\sum_{i\neq j}t_{ij}^{\perp}c_i^{\dag}c_j + \sum_{i}U_ic_i^{\dag}c_i,
\end{equation}
where the summation runs over all lattice sites of the system,
$\varepsilon_i$ is the on-site energy at site $i$, $t_{ij}^{\|}$ ($t_{ij}^{\perp}$) is the intralayer (interlayer) hopping energy between sites $i$ and $j$, $U_i$ is the electrostatic potential energy at site $i$, and $c_i^{\dag}$ ($c_j$) is the creation (annihilation) operator of an electron at site $i$ ($j$). For simplicity, the on-site energy $\varepsilon_i$ is set to zero for all the lattice sites. It was shown \cite{Rudenkorealisticdescriptionmultilayer2015} that with ten intralayer and five interlayer hopping parameters, this TB model can well describe the band structure of multilayer phosphorene in the low-energy region when compared to that obtained by DFT-GW calculations. The ten intralayer hopping parameters (in units of eV) are $t_{1}^{\|}=-1.486$, $t_{2}^{\|}=+3.729$, $t_{3}^{\|}=-0.252$, $t_{4}^{\|}=-0.071$, $t_{5}^{\|}=+0.019$, $t_{6}^{\|}=+0.186$, $t_{7}^{\|}=-0.063$, $t_{8}^{\|}=+0.101$, $t_{9}^{\|}=-0.042$, $t_{10}^{\|}=+0.073$, and the five interlayer hopping parameters (in units of eV) are $t_{1}^{\perp}=+0.524$, $t_{2}^{\perp}=+0.180$, $t_{3}^{\perp}=-0.123$, $t_{4}^{\perp}=-0.168$, $t_{5}^{\perp}=+0.005$ \cite{Rudenkorealisticdescriptionmultilayer2015}. These hopping parameters are illustrated in Figs. \ref{fig1}(c) and \ref{fig1}(d).
In the following, we show how to obtain the electrostatic potential energy $U$ within a self-consistent Hartree approximation.

We consider undoped multilayer phosphorene in the presence of an electrostatic gating. This situation can be realized in an experimental setup with external top and bottom gates having opposite voltages applied to multilayer phosphorene \cite{DengEfficientelectricalcontrol2017}. As shown in Fig. \ref{fig1}, a positively (negatively) charged gate is placed on the top (bottom) of multilayer phosphorene of $N$ layers ($N=3$ in the figure). The top (bottom) gate is assumed to have a positive (negative) charge density $n_t>0$ ($n_b<0$) on it. These two gates are used to generate and control the carrier densities in the system. Because a single layer of phosphorene can be viewed as consisting of two atomic sublayers due to its puckered lattice structure, both of these two sublayers can be treated as \textit{individual} layers and thus the number of \textit{individual} layers of $N$-layer phosphorene is $2N$. This is significantly different from multilayer graphene, where the number of \textit{individual} layers is just $N$ for $N$-layer graphene. Therefore, in the $N$-layer phosphorene system a total carrier density $n=n_t+n_b=\sum_{i}n_i$ is induced, where $n_i$ is the carrier density on the $i$-th sublayer and the summation is over all the $2N$ sublayers.

In our model, we assume that the top (bottom) gate produces a uniform electric field $F_t=en_t/(2\varepsilon_0\kappa)$ [$F_b=en_b/(2\varepsilon_0\kappa)$], which can be obtained from fundamental electrostatics, where $e$ is the elementary charge, $\varepsilon_0$ is the permittivity of vacuum, and $\kappa$ is the dielectric constant. The induced charge carriers on the phosphorene sublayers, in its turn, produce a uniform electric field $F_i=n_ie/(2\varepsilon_0\kappa)$ ($i=1, 2, ..., 2N$), which counteracts the electric field produced by the external gates. The inversion asymmetry between the two adjacent sublayers $i$ and $i+1$ is determined by a potential energy difference $\Delta_{i,i+1}$, which is given by
\begin{equation}\label{e2}
\Delta_{i,i+1}= \alpha_{i,i+1}\big(\sum_{j=i+1}^{2N}n_j+|n_b|\big),
\end{equation}
where $\alpha_{i,i+1}=e^2d_{i,i+1}/(\varepsilon_0\kappa)$ with $d_{i,i+1}$ the distance between the two adjacent sublayers. In multilayer phosphorene, this inter-sublayer distance is not constant due to the puckered lattice structure of phosphorene: it is $d_{i, i+1}=d_{intra}$ if the sublayers $i$ and $i+1$ are within the same phosphorene layer while it is $d_{i, i+1}=d_{inter}$ if these two sublayers belong to different phosphorene layers, with $d_{intra}=0.212$ nm and $d_{inter}=0.312$ nm. The total electric field between the two adjacent sublayers is given by $F_{i,i+1}=\Delta_{i,i+1}/d_{i,i+1}$ ($i=1, 2, ..., 2N-1$). Finally, the electrostatic Hartree energies $U_i$, which are added to the $i$-th sublayer on-site elements of the $N$-layer TB Hamiltonian \eqref{e1}, can be obtained as
\begin{equation}\label{e3}
U_i=0 \ (i=1); \ U_i=\sum_{j=1}^{i-1}\Delta_{j,j+1} \ (i>1)
\end{equation}
Here we assumed zero electrostatic potential energy on the top-most sublayer (i.e., the sublayer that is closest to the top gate).

Due to the in-plane translational invariance of the system, a Fourier transform is performed to convert the $N$-layer TB Hamiltonian \eqref{e1} into momentum space, and then the converted Hamiltonian is numerically diagonalized to obtain the eigenvalues and eigenvectors of the system. All numerical calculations are performed using the recently developed TB package PYBINDING \cite{MoldovanPybinding2016}. Because there are four inequivalent basis atoms (labeled as $A$, $B$, $C$, and $D$) in an unit cell of phosphorene, the dimension of the Fourier-transformed TB Hamiltonian of the $N$-layer phosphorene system is $4N \times 4N$. Therefore, the corresponding eigenvectors are the column vectors of dimension $4N$ consisting of the coefficients of the TB wave functions,
\begin{equation}\label{e4}
c=[c_{A_1}, c_{B_1}, c_{C_1}, c_{D_1}, ..., c_{A_N}, c_{B_N}, c_{c_N}, c_{D_N}]^T,
\end{equation}
where $c_{A_i}$, $c_{B_i}$, $c_{C_i}$, $c_{D_i}$ are the $i$-th layer coefficients for basis atoms $A$, $B$, $C$, $D$, respectively, and the symbol $T$ denotes the transpose of a vector or matrix. Note that these TB coefficients depend on the in-plane wave vector $\textbf{k}$. The total TB wave function of the $N$-layer phosphorene system is then given by
\begin{equation}\label{e5}
\Psi=\sum_{i=1}^N\big[c_{A_i}\psi_{A_i}+c_{B_i}\psi_{B_i}+ c_{C_i}\psi_{C_i} + c_{D_i}\psi_{D_i}\big],
\end{equation}
where $\psi_{A_i}$, $\psi_{B_i}$, $\psi_{C_i}$, $\psi_{D_i}$ are the four components of the $i$-th layer TB wave function. With the obtained layer-dependent coefficients $c_i=[c_{A_i}, c_{B_i}, c_{C_i}, c_{D_i}]$, the carrier densities on the sublayers of phosphorene are given by
\begin{equation}\label{e6}
\begin{aligned}
n_{2i-1} = 2\sum_{\textbf{k}}f[E(\textbf{k})](|c_{A_i}|^2+|c_{D_i}|^2), \\
n_{2i} = 2\sum_{\textbf{k}}f[E(\textbf{k})](|c_{B_i}|^2+|c_{C_i}|^2),
\end{aligned}
\end{equation}
where $i=1, 2, ..., N$, the factor 2 in front of the summations accounts for the spin degeneracy, $E(\textbf{k})$ is the energy spectrum obtained by numerically diagonalizing the TB Hamiltonian \eqref{e1} in momentum space, and $f[E(\textbf{k})]$ is the Fermi-Dirac function describing the carrier distribution in the energy spectrum. In the presence of electrical gating, the carrier densities in fully occupied energy bands are changed and one has to take into account the charge density redistribution in the valence bands.

The TB Hamiltonian \eqref{e1} depends on the gate-induced carrier densities through Eqs. \eqref{e2} and \eqref{e3}, which in turn are calculated based on the full eigenstates of the TB Hamiltonian \eqref{e1} in momentum space. Therefore, a self-consistent calculation using Eqs. \eqref{e1}, \eqref{e2}, \eqref{e3} and \eqref{e6} is required to obtain the carrier densities $n_i$ and the Hartree energies $U_i$ on the different sublayers of phosphorene. Since the carrier densities on the different sublayers are not known in advance, an initial guess of these densities is needed, e.g., assuming them to be equal at the beginning. Then the calculations are performed self-consistently until the carrier density per sublayer is converged. The number of self-consistent iterations depends on the total carrier density $n$ and on the number of stacking layers $N$. When convergency is reached, the Fermi energy $E_F$ and band structure $E_\textbf{k}$ of multilayer phosphorene can be obtained, from which we can further calculate the electronic properties such as the fundamental band gap and the carrier effective mass.

\begin{figure}[htb!]
\begin{center}
\includegraphics[width=0.45\textwidth]{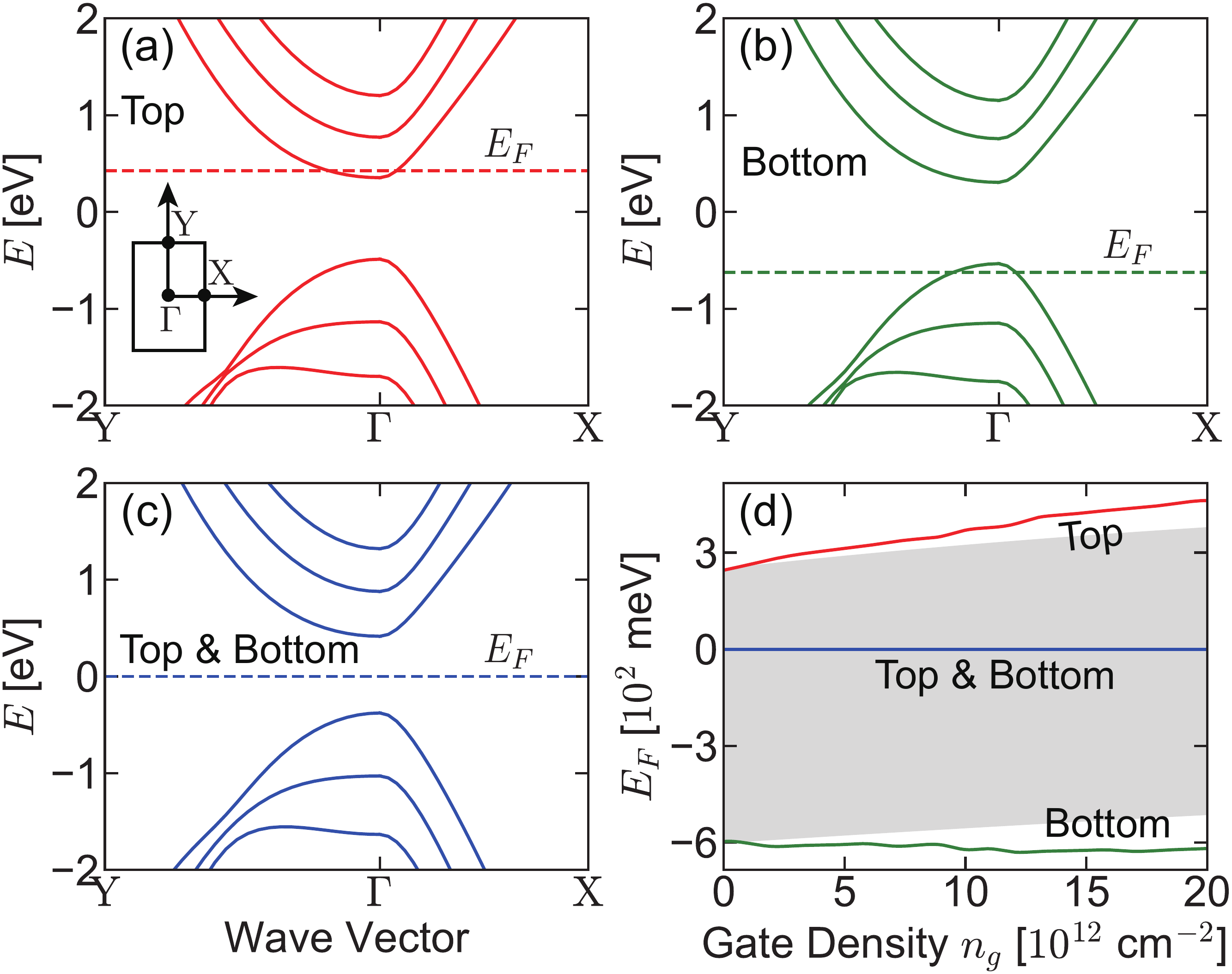}
\caption{(a)-(c) Band structure of gated multilayer phosphorene obtained by the self-consistent TB model: (a) in the presence of only top gate ($n_t=1.5\times10^{13}$ cm$^{-2}$), (b) in the presence of only bottom gate ($n_b=-1.5\times10^{13}$ cm$^{-2}$), and (c) in the presence of both top and bottom gates ($n_t=-n_b=1.5\times10^{13}$ cm$^{-2}$). (d) Fermi energy $E_F$ as a function of the gate charge density $n_g$ for top gating ($n_g=n_t$), bottom gating ($n_g=n_b$), and both top and bottom gating ($n_g=n_t=-n_b$). In panels (a)-(c), the horizontal dashed lines are the self-consistently determined Fermi energies. The inset in (a) denotes the first Brillouin zone (BZ) of phosphorene, where $\Gamma$, X and Y are the three most important high-symmetry points. The shaded region in (d) represents the band-gap region.}\label{fig2}
\end{center}
\end{figure}

\section{Results and Discussion}

In the present work, we consider three different gate configurations applied to multilayer phosphorene: (i) only positive top gate $V_t>0$, $V_b=0$ ($n_t>0$, $n_b=0$); (ii) only negative bottom gate $V_t=0$, $V_b<0$ ($n_t=0$, $n_b<0$); and (iii) both top and bottom gates $V_t=-V_b>0$ ($n_t=-n_b>0$). We calculate self-consistently the band structure of gated multilayer phosphorene and the gate-induced charge densities on the different phosphorene layers. In our self-consistent calculations, we varied the number of phosphorene layers $N$ and set the temperature to $T=10$ K. The charge density $n_t$ ($n_b$) on the top (bottom) gate was changed between 0 and $2\times10^{13}$ cm$^{-2}$. By using the simple relation between the gate voltage potential $V_g$ and the gate charge density $n_g$ \cite{AvetisyanElectricfieldcontrolband2009}, $V_{g}=en_{g}d/(2\varepsilon_0\varepsilon)$ with $d$ ($\varepsilon$) the thickness (relative permittivity) of the dielectric material, and by taking the experimental values of $d$ and $\varepsilon$ \cite{DengEfficientelectricalcontrol2017}, the chosen charge-density range for the top (bottom) gate was found to correspond to the top (bottom) gate voltage between 0 and 10 V (40 V) that are experimentally accessible \cite{DengEfficientelectricalcontrol2017}. Furthermore, in our numerical calculations we took the dielectric constant $\kappa=6$ for multilayer phosphorene. We chosen this $\kappa$ value because (i) it is close to the dielectric constant of bulk BP ($\kappa=8.3$), (ii) it is within the range of the $\kappa$ value of multilayer phosphorene determined by DFT calculations \cite{KumarThicknesselectricfielddependentpolarizability2016a}, (iii) there are not yet experimental $\kappa$ values reported for multilayer phosphorene, and (iv) it leads to a qualitatively good agreement between theory and experiment (as will be shown later).

\begin{figure*}[htb!]
\begin{center}
\includegraphics[width=0.8\textwidth]{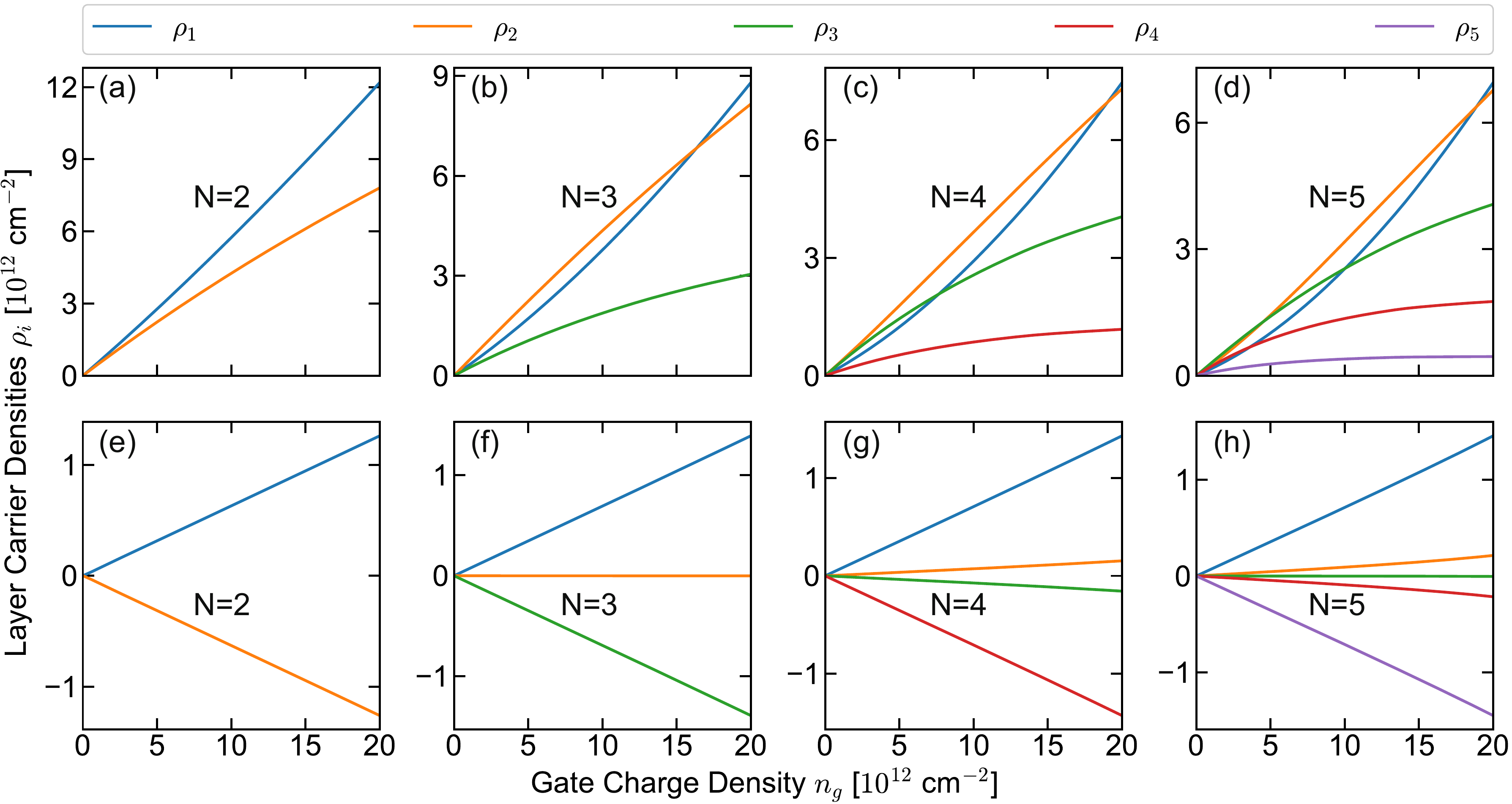}
\caption{Layer carrier densities $\rho_i$ ($i=1, ..., N$) of gated multilayer phosphorene as a function of the gate charge density $n_g$ for $N=2, 3, 4, 5$ layers: (a)-(d) in the presence of only a top gate ($n_g=n_t$), and (e)-(h) in the presence of both top and bottom gates ($n_g=n_t=-n_b$).}\label{fig3}
\end{center}
\end{figure*}

\begin{figure}[htb!]
\begin{center}
\includegraphics[width=0.45\textwidth]{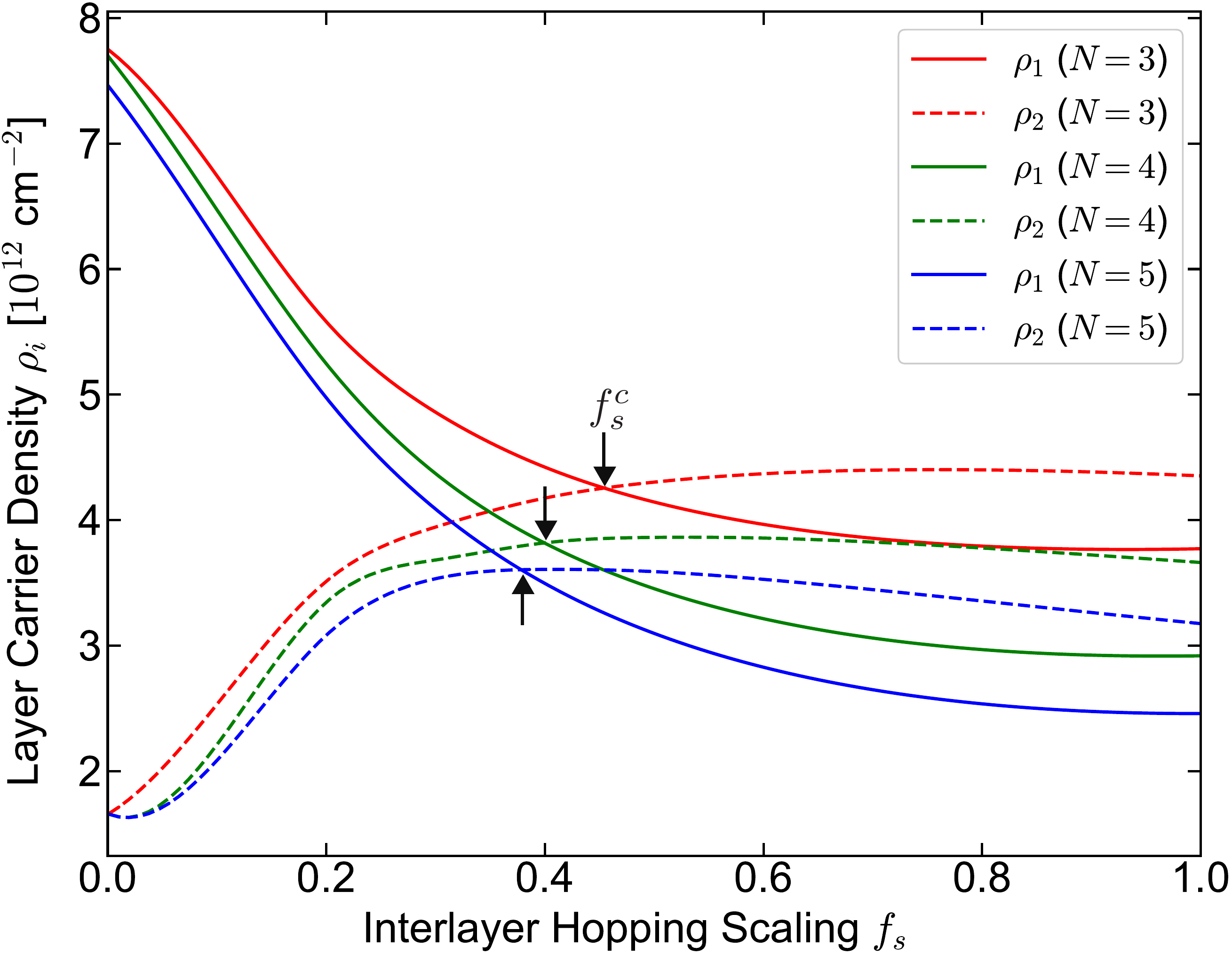}
\caption{Layer carrier densities $\rho_{1}$ and $\rho_2$ of top-gated multilayer phosphorene as a function of the interlayer hopping scaling factor $f_s$ for $N=3, 4, 5$ layers. The short arrow indicates the critical value ($f_s^c$) of the interlayer hopping strength, below (above) which $\rho_1>\rho_2$ ($\rho_1<\rho_2$) is satisfied.}\label{fig4}
\end{center}
\end{figure}

\subsection{Band Structure and Fermi Energy}

In Figs. \ref{fig2}(a)-(c), we show the self-consistently obtained band structure and Fermi energy of three-layer phosphorene in the presence of (a) only a top gate, (b) only a bottom gate, and (c) both top and bottom gates. The gate charge densities are assumed to be $n_t=10^{13}$ cm$^{-2}$ in (a), $n_b=-10^{13}$ cm$^{-2}$ in (b), and $n_t=-n_b=10^{13}$ cm$^{-2}$ in (c). As can be seen, in the presence of only a top (bottom) gate, the Fermi energy is located in the conduction (valence) band due to $n_t>0$ ($n_b<0$), which indicates a finite density of electrons (holes) in the system; whereas in the presence of both top and bottom gates, the Fermi energy is located in the band gap due to $n_t+n_b=0$, which indicates no excess carriers in the system.

In Fig. \ref{fig2}(d), we show the dependence of the Fermi energy ($E_F$) of three-layer phosphorene on the gate charge density ($n_g$) for the cases of only a top gate ($n_g=n_t$), only a bottom gate ($n_g=n_b$), and in the presence of both top and bottom gates ($n_g=n_t=-n_b$). The shaded region in this figure represents the band-gap region. As can be seen, applying only a top (bottom) gate enables to tune the Fermi energy of the system into the conduction (valence) band region and so the electron (hole) density of the system can be tuned with varying gate charge density. However, with both top and bottom gates applied, it is possible to tune the Fermi energy of the system into the band-gap region with no net carrier density when varying gate charge density. Notice that in the presence of only a top or bottom gate, the Fermi energy variation with the gate charge density is smaller in three-layer phosphorene than that in three-layer graphene \cite{AvetisyanElectricfieldtuning2009,
AvetisyanElectricfieldcontrolband2009}. This is due to the fact that the carrier effective mass of multilayer phosphorene is larger than that of multilayer graphene. We also observe similar results for other multilayer phosphorene (e.g., two-, four- and five-layer phosphorene).

\begin{figure*}[htb!]
\begin{center}
\includegraphics[width=0.8\textwidth]{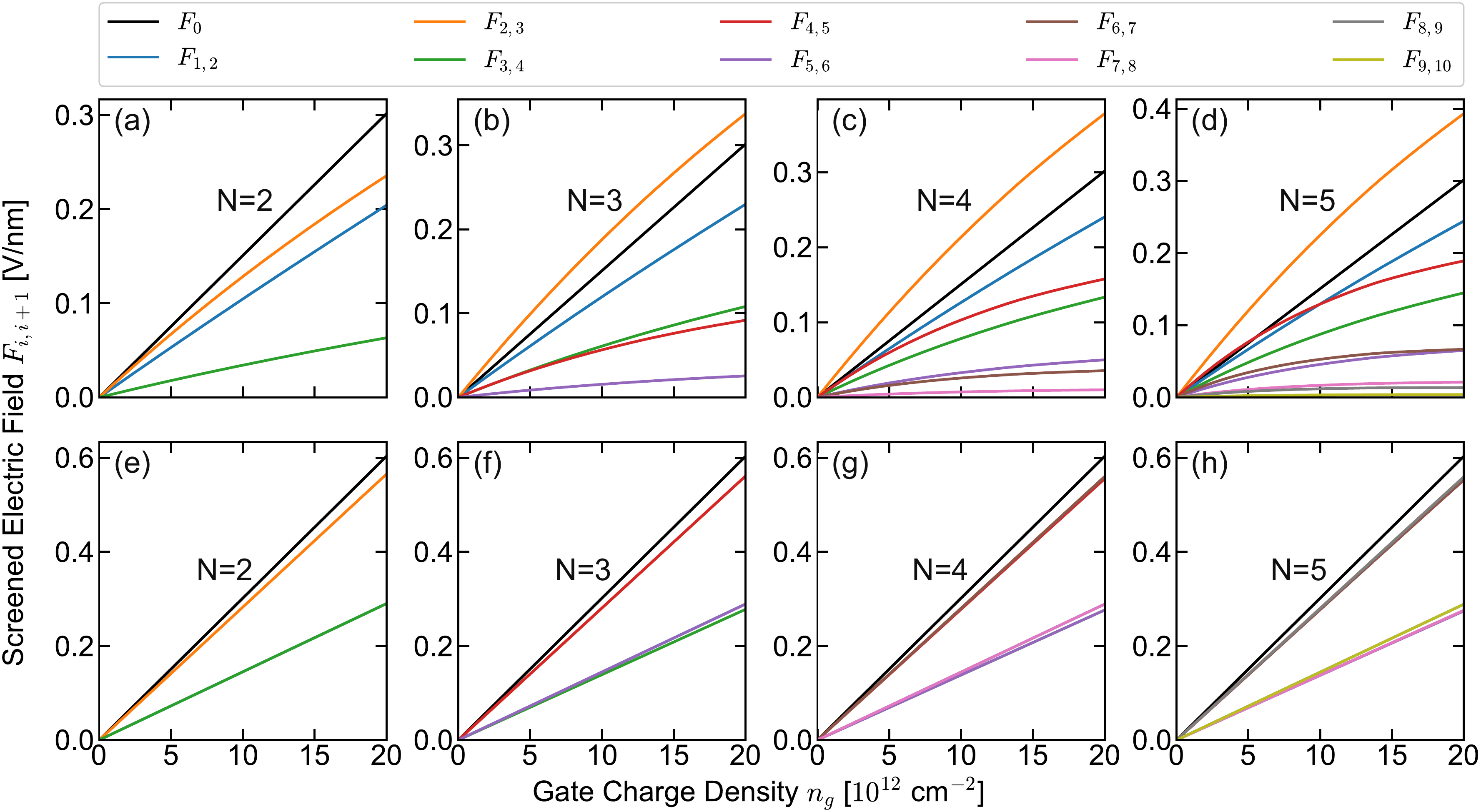}
\caption{Electric fields $F_{i, i+1}$ ($i=1, ..., 2N-1$) between the two adjacent sublayers of gated multilayer phosphorene as a function of the gate charge density $n_g$ for $N=2, 3, 4, 5$ layers: (a)-(d) in the presence of only a top gate ($n_g=n_t$), and (e)-(h) in the presence of both top and bottom gates ($n_g=n_t=-n_b$). In each panel, $F_0$ (black curve) is the electric field produced by top and/or bottom gates.}\label{fig5}
\end{center}
\end{figure*}

\subsection{Charge Distribution and Screening}

With top and/or bottom gates applied to multilayer phosphorene, the gate-produced electric field induces a charge distribution over the phosphorene layers, which in turn produces an internal electric field between the layers that counteracts the external (gate-produced) electric field (i.e., the electric-field-induced charge screening). Although the gate-produced (unscreened) electric field is uniform, the screened electric field between the different phosphorene layers is expected to be not uniform, because the induced charge densities on these layers are generally not equal to each other.

In Fig. \ref{fig3}, we show the carrier densities $\rho_i$ ($i=1, ..., N$) on the different layers of $N$-layer phosphorene ($N=2, 3, 4, 5$) as a function of the gate charge density $n_g$: (a)-(d) in the presence of only a top gate ($n_g=n_t$), and (e)-(h) in the presence of both top and bottom gates ($n_g=n_t=-n_b)$. The relation between the layer and sublayer densities is given by $\rho_i=n_{2i-1}+n_{2i}$. The results are very similar in the presence of only a bottom gate ($n_g=-n_b)$, and can be mapped into each other by reversing the layer index and changing the carrier type, and thereby we do not show them.

We see that in the presence of only a top gate, as shown in Figs. \ref{fig3}(a)-\ref{fig3}(d), the top-most layer has the largest carrier density (i.e., $\rho_1$) for the case of $N=2$ layers. But this is no longer true for the cases of $N=3, 4, 5$ layers. For instance, for $N=4$ the carrier density is largest on the second-top-most layer (i.e., $\rho_2$). This carrier-density anomaly in $N$-layer phosphorene (with $N\geq 3$) is induced by the charge transfer between the different layers due to the significant interlayer coupling. Our numerical calculations show that if the strength of the interlayer coupling is reduced below a critical value, the top-most phosphorene layer will regain the largest carrier density. In order to see this more clearly, we show in Fig. \ref{fig4} the carrier densities on the top-most and the second-top-most phosphorene layers ($\rho_1$ and $\rho_2$) as a function of the interlayer hopping scaling factor ($f_s$) for the cases of $N=3, 4, 5$ layers, where $f_s=0$ ($1$) indicates the turning off (on) of the full interlayer hopping. As can be seen, there is indeed a critical value of the interlayer hopping strength (indicated by $f_s^c$ in the figure), below (above) which we have $\rho_1>\rho_2$ ($\rho_1<\rho_2$), and the value of $f_s^c$ depends on the number of layers $N$ (for instance it is found to be $f_s^c=0.45, 0.39, 0.37$ for $N=3, 4, 5$). This indicates the importance of the interlayer coupling strength in the determination of the carrier densities over the different phosphorene layers.

Furthermore, we find that in the presence of both top and bottom gates, as shown in Figs. \ref{fig3}(e)-\ref{fig3}(h), the carrier densities on the different layers exhibit an odd-even layer dependence: for the case of an even number of layers ($N=2, 4$), the upper and lower layers with respect to the centro-symmetric plane have carrier densities opposite in sign but equal in magnitude, thereby leading to the appearance of electron-hole bilayers; whereas for the case of an odd number of layers ($N=3, 5$), there is an additional feature that the middle layer (at the centro-symmetric plane) has zero carrier density. The layer carrier densities $\rho_i$ increase with the gate charge density $n_g$ in either a linear or nonlinear fashion.

\begin{figure}[htb!]
\begin{center}
\includegraphics[width=0.45\textwidth]{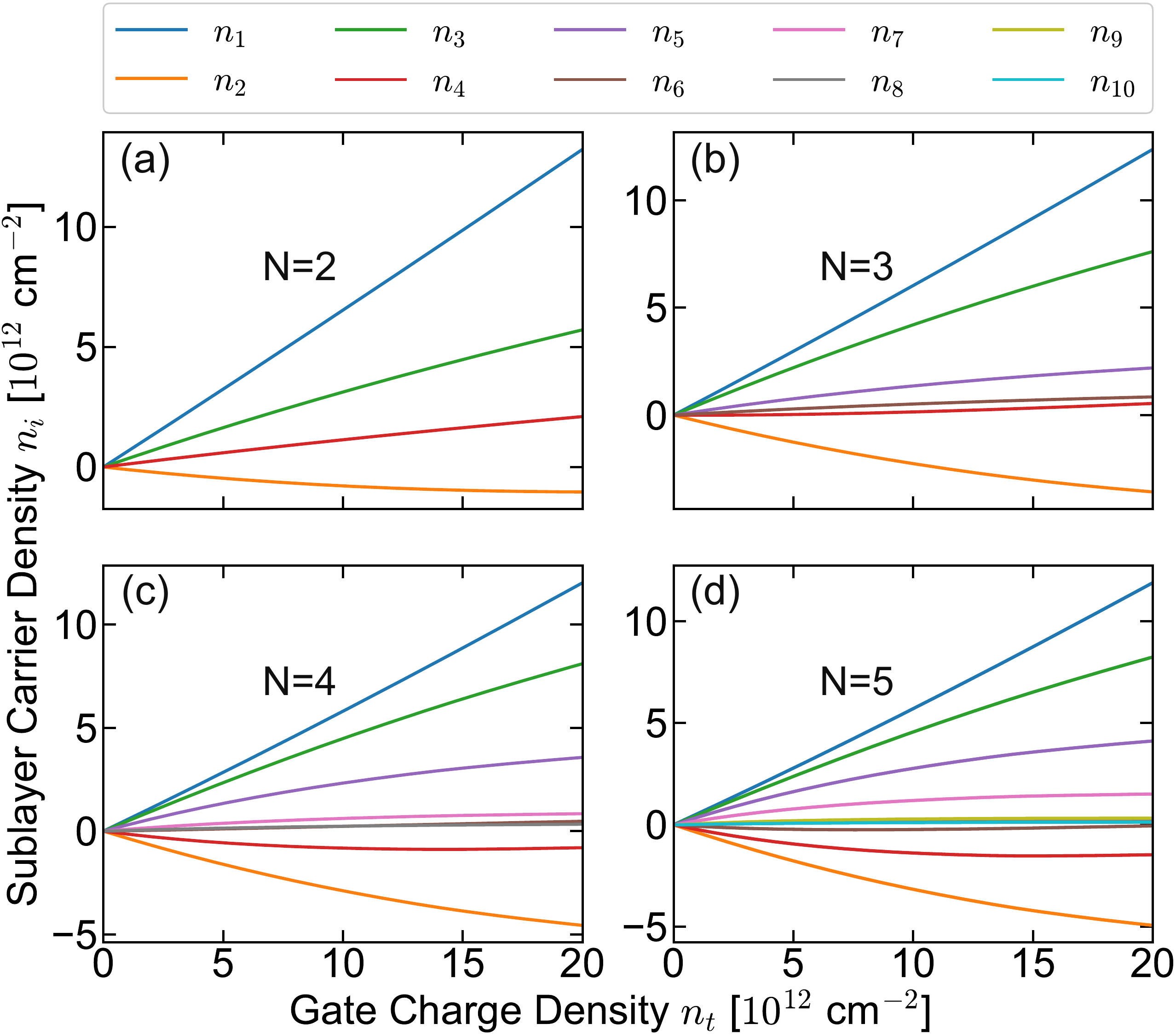}
\caption{Sublayer carrier densities $n_{i}$ ($i=1, ..., 2N$) of top-gated multilayer phosphorene as a function of the gate charge density $n_t$ for $N=2, 3, 4, 5$ layers.}\label{fig6}
\end{center}
\end{figure}

In Fig. \ref{fig5}, we show the total electric fields $F_{i,i+1}$ ($i=1, ..., 2N-1$) between the two adjacent sublayers of $N$-layer phosphorene ($N=2, 3, 4, 5$) as a function of the gate charge density $n_g$, where (a)-(d) and (e)-(h) are for the same gating configurations as in Fig. \ref{fig3}. For comparative purposes, the gate-produced (unscreened) electric field $F_0=F_t+F_b$ is also presented in each panel. As can be seen, in all the gating configurations, most of the electric fields $F_{i, i+1}$ are significantly smaller as compared to the fields $F_0$ due to the charge screening effect, and the magnitudes of $F_{i, i+1}$ are different from each other, because the carrier densities on the different sublayers that can screen the electric field $F_0$ are different (see Fig. \ref{fig6}). Therefore, the screened electric field across multilayer phosphorene is not uniform, whose magnitude depends on the phosphorene sublayers.

Notice that in the presence of only a top or bottom gate and for the cases of $N\geq3$ layers, some electric fields $F_{i, i+1}$ shown in Fig. \ref{fig5}, e.g., $F_{2,3}$, $F_{4,5}$ and $F_{6,7}$, are not screened and even larger than the unscreened one $F_0$. This charge-screening anomaly is induced by the fact that there are emerging minority carriers (with opposite sign to the majority ones) on the sublayers in the presence of only a top or bottom gate. For instance, for the case of $N=4$ layers and in the presence of only top gate, as shown in Fig. \ref{fig6}(c), they are the carrier densities $n_2$ and $n_4$ on the second- and fourth-top-most sublayers, which are opposite in sign as compared to the other densities. With increasing $N$, the minority carriers appear on more sublayers. The emergence of minority carriers is induced by the intralayer charge transfer due to the very strong out-of-plane hopping strength (i.e., $t_2^{\|}$) in phosphorene. The emerging minority carriers produce an electric field, which counteracts that produced by the majority carriers and thus leads to the charge-screening anomaly.

We also note that in the presence of only a top gate, as shown in Figs. \ref{fig5}(a)-\ref{fig5}(d), some of the electric fields $F_{i, i+1}$ exhibit a nonlinear screening with respect to the gate charge density (or gate electric field) while others show a linear screening. However, this is different in the presence of both top and bottom gates, as shown in Figs. \ref{fig5}(e)-\ref{fig5}(h), where all the electric fields $F_{i, i+1}$ show a linear screening with gate charge density. This difference occurs due to the either linear or nonlinear increase of the layer carrier densities with the gate charge density, as shown in Fig. \ref{fig3}. Moreover, in the presence of dual gates the screened electric fields $F_{i, i+1}$ are symmetric with respect to the centro-symmetric plane due to the symmetric carrier distribution in this gating configuration.

It is worth to point out that in previous studies of multilayer graphene \cite{McCannAsymmetrygapelectronic2006,AvetisyanElectricfieldtuning2009,
AvetisyanElectricfieldcontrolband2009,KoshinoGateinducedinterlayerasymmetry2009,
ZhangBandstructureABC2010}, the charge screening induced by the gate electric field is only present between the graphene layers (i.e., interlayer screening). Here, we find that it can also be present within the phosphorene layer (i.e., intralayer screening), which arises due to the puckered lattice structure of phosphorene. Therefore, both intralayer and interlayer charge screenings are present in multilayer phosphorene, which is fundamentally different from multilayer graphene.

It should also be noted that the present TB model takes into account only one atomic orbital (i.e., the $p_z$ orbital) per lattice site and thus only the electrons on the $p_z$ orbital contribute to the charge screening effect. When more atomic orbitals (e.g., $s$, $p_x$, $p_y$ orbitals) are included in the model, the electrons on these orbitals may also contribute to the charge screening effect. Therefore, the charge screening effect is probably underestimated by the current TB model. However, it was shown in Ref. \cite{Rudenkorealisticdescriptionmultilayer2015} that when compared to the DFT-GW approach, this single-orbital TB model reproduces well the electronic band structure of multilayer phosphorene in the low-energy region and (ii) the $p_z$ orbital that was included in the model has the predominant contribution to the electronic band structure in the low-energy region. Therefore, the present TB model is expected to be qualitatively accurate in describing the charge screening effect in multilayer phosphorene. With the inclusion of more atomic orbitals, the magnitude of charge screening could be changed, which however might not change the main conclusions on the charge screening effect obtained by the present TB model.

\subsection{Effective Mass and Band gap}

\begin{figure}[htb!]
\begin{center}
\includegraphics[width=0.45\textwidth]{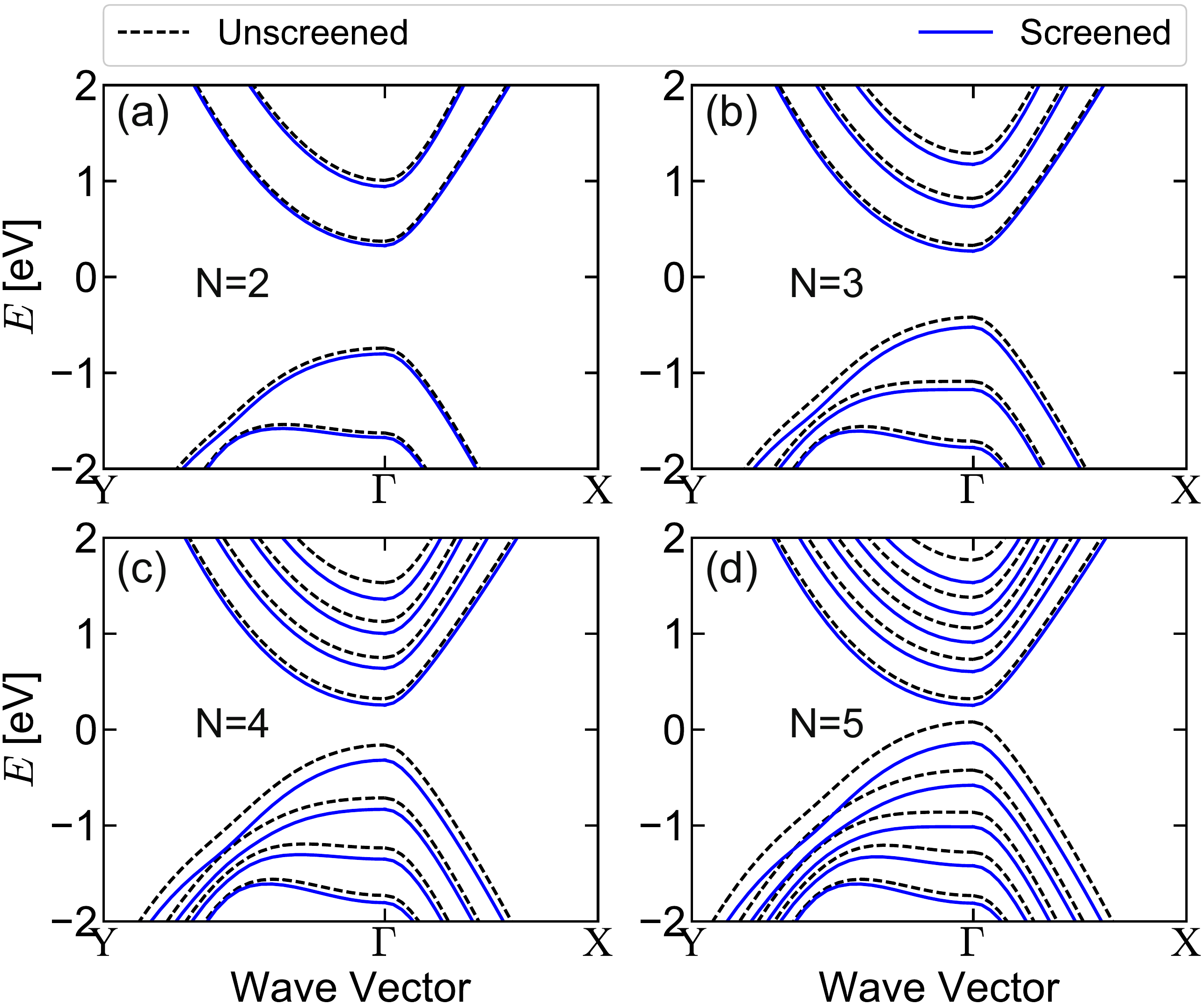}
\caption{Screened (full curves) and unscreened (dashed curves) band structures of multilayer phosphorene in the presence of both top and bottom gates with fixed charge densities of $n_t=-n_b=1.5\times10^{13}$ cm$^{-2}$ for $N=2, 3, 4, 5$ layers.}\label{fig7}
\end{center}
\end{figure}

Now we turn to the effect of charge screening on the gate-electric-field tuning of the effective mass and the band gap of multilayer phosphorene. To proceed, we first show how the band structure of gated multilayer phosphorene is different in the absence and presence of charge screening. In Fig. \ref{fig7}, we plot both the screened and unscreened band structures of $N$-layer phosphorene in the presence of both top and bottom gates with fixed charge densities ($n_t=-n_b=1.5\times10^{13}$ cm$^{-2}$), where (a)-(d) are for $N=2$-$5$ layers, respectively. It can be clearly seen from this figure that in the presence of charge screening, the band structure is significantly changed, and that the change is more pronounced for the case of larger number of layers.

\begin{figure}[htb!]
\begin{center}
\includegraphics[width=0.45\textwidth]{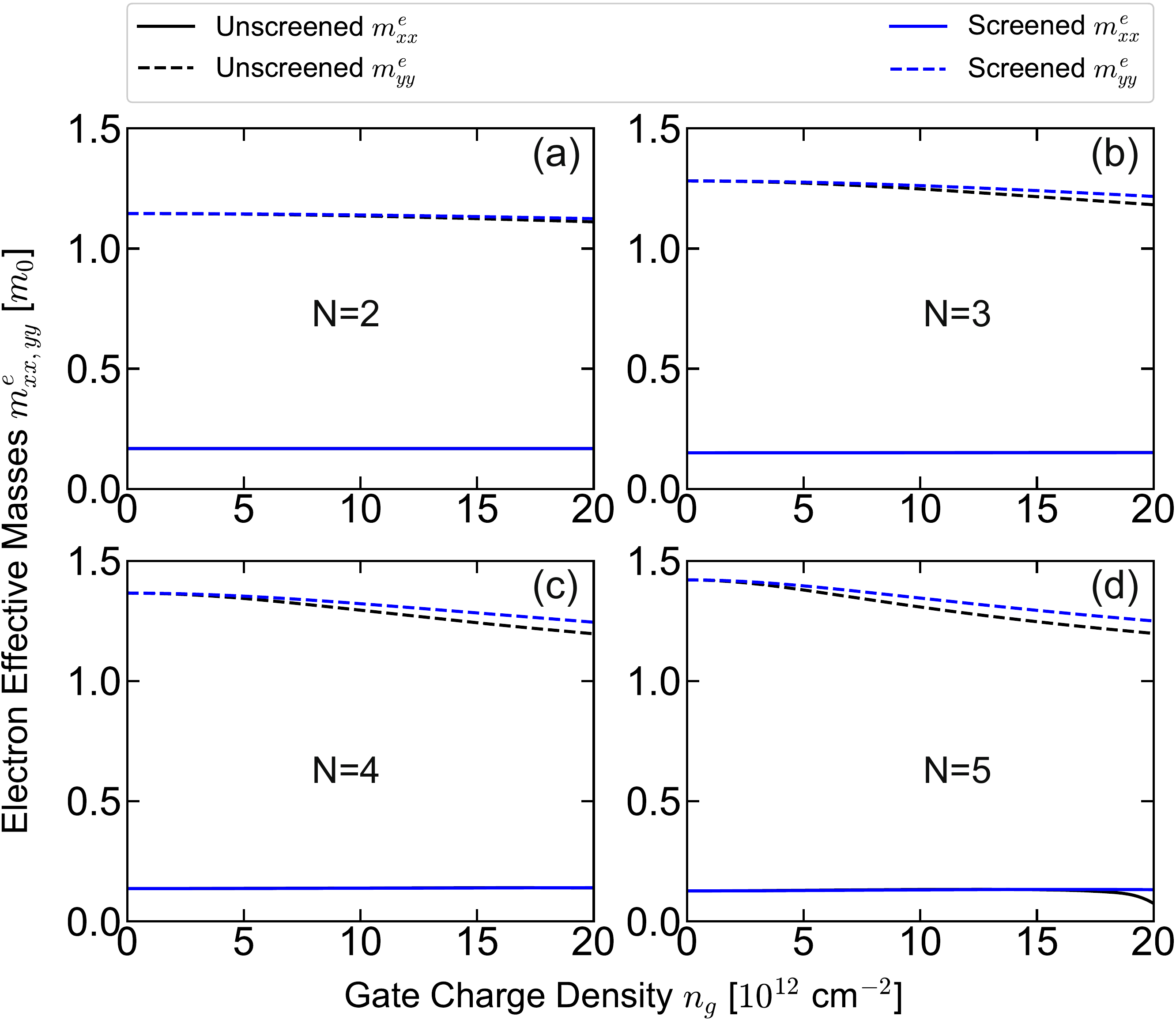}
\caption{Screened (full curves) and unscreened (dashed curves) electron effective masses of multilayer phosphorene in the presence of both top and bottom gates with varying gate charge density for $N=2, 3, 4, 5$ layers, where $m_0$ is the free electron mass. }\label{fig8}
\end{center}
\end{figure}

From the electronic band structure $E(\textbf{k})$, the carrier (electron or hole) effective mass can be calculated by $1/m_{ij}=\partial^2 E(\textbf{k})/(\hbar^2\partial k_i\partial k_j)$, with $i, j= x, y$ and $m_{ij}$ the effective mass tensor. In Fig. \ref{fig8}, we show the electron effective masses of $N$-layer phosphorene along the armchair and zigzag directions ($m_{xx}^e$ and $m_{yy}^e$) with varying gate charge density ($n_g$) in the presence of both top and bottom gates ($n_g=n_t=-n_b$), where (a)-(d) are for $N=2$-$5$ layers, respectively. As can be seen, the electron effective mass along the armchair direction $m_{xx}^e$ changes more significantly with varying gate charge density $n_g$ than that along the zigzag direction $m_{yy}^e$. This is primarily due to the band anisotropy of multilayer phosphorene. In addition, the gate-induced charge screening has different consequences on the electron effective masses $m_{xx}^e$ and $m_{yy}^e$: it has a more significant effect on $m_{yy}^e$ than $m_{xx}^e$. However, the variations of both the electron effective masses $m_{xx}^e$ and $m_{yy}^e$ with gate charge density $n_g$ are not significantly influenced by the gate-induced charge screening, as can be verified by the band dispersion curvatures shown in Fig. \ref{fig7}. Similar results are also obtained for the hole effective masses $m_{xx}^h$ and $m_{yy}^h$.

\begin{figure}[htb!]
\begin{center}
\includegraphics[width=0.45\textwidth]{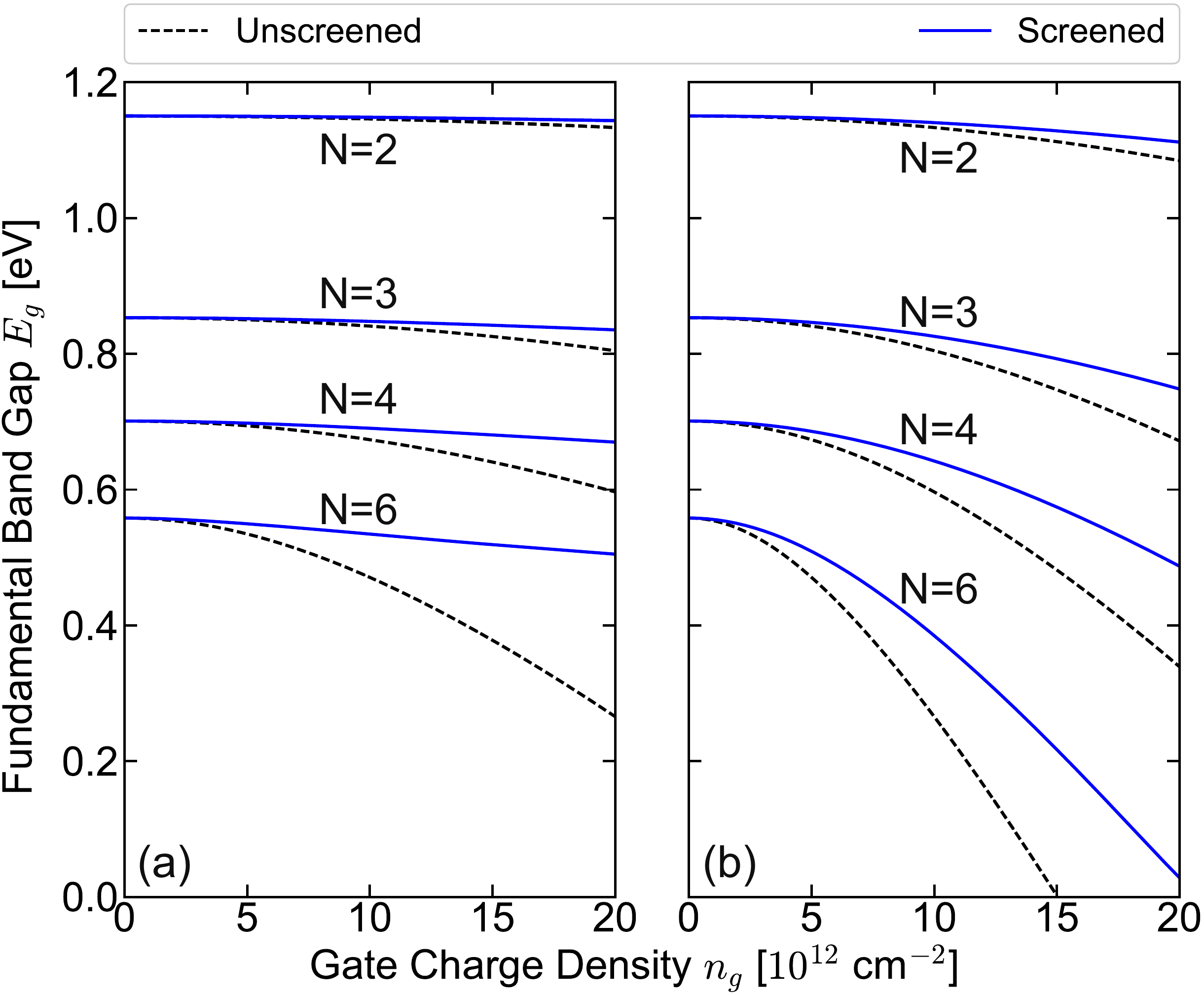}
\caption{Screened (full curves) and unscreened (dashed curves) band gaps of multilayer phosphorene as a function of the gate charge density $n_g$ for $N=2, 3, 4, 6$ layers: (a) in the presence of only a top gate ($n_g=n_t$), and (b) in the presence of both top and bottom gates ($n_g=n_t=-n_b$) .}\label{fig9}
\end{center}
\end{figure}

\begin{table*}
\caption{The Fitting parameters $a$, $b$ and $c$ for the gate charge density ($n_g$ in units of cm$^{-2}$) dependence of the fundamental band gap ($E_g$ in units of eV) $E_g=an_g^2+bn_g+c$ for $N$-layer phosphorene ($N=2, 3, 4, 6$) in the presence of both top and bottom gates ($n_g=n_t=-n_b$), where $a$, $b$ and $c$ are in units of eV$\cdot$cm$^{2}$, eV$\cdot$cm and eV, respectively.}\label{tab1}
\begin{ruledtabular}
\begin{tabular}{cccccc}
Model             & Parameters & $N=2$ & $N=3$ & $N=4$ & $N=6$ \\
\hline
                 & $a$ & $-1.583\times10^{-4}$ & $-4.219\times10^{-4}$ & $-7.663\times10^{-4}$ & $-1.741\times10^{-3}$ \\

Unscreened       & $b$ & $-1.541\times10^{-4}$ &
                 $-7.772\times10^{-4}$
                 & $-3.268\times10^{-3}$ & $-1.243\times10^{-2}$ \\

                & $c$ & $1.151$ & $ 0.855$  & 0.704 & $0.571$ \\
\hline
                & $a$ & $-9.422\times10^{-5}$ & $-2.517\times10^{-4}$ & $-4.815\times10^{-4}$ & $-9.177\times10^{-4}$ \\

Screened        & $b$ & $-5.315\times10^{-5}$ &
                $-2.722\times10^{-4}$
                & $-1.277\times10^{-3}$ & $-9.415\times10^{-3}$ \\

                & $c$ & $1.151$ & $0.855$  & $0.704$ & $0.571$ \\
\end{tabular}
\end{ruledtabular}
\end{table*}

In Fig. \ref{fig9}, we show both the screened and unscreened band gaps ($E_g$) of $N$-layer phosphorene ($N=2, 3, 4, 6$) as a function of the gate charge density ($n_g$) in the presence of (a) only a top gate ($n_g=n_t$) and (b) both top and bottom gates ($n_g=n_t=-n_b$). As can be seen, the gate-charge-density (or gate-electric-field) tuning of the band gap is distinctively different in the absence and presence of charge screening. The unscreened band gap decreases dramatically with increasing gate charge density (or gate electric field). However, in the presence of charge screening, the magnitude of this band-gap decrease is significantly reduced and the reduction is more significant for larger number of layers. For instance, for the case of $N=6$ layers and in the presence of both top and bottom gates, the unscreened band gap $E_g$ becomes zero when the gate charge density $n_g$ reaches the critical value $n_g^c\sim1.5\times10^{13}$ cm$^{-2}$; however, due to the charge screening, the critical gate charge density $n_g^c$ is significantly increased, which becomes larger than $2\times10^{13}$ cm$^{-2}$ and thus increases by more than $33\%$. The effect of the charge screening on the band-gap tuning with the gate electric field is qualitatively the same for all the gating configurations, as shown in Figs. \ref{fig9}(a) and \ref{fig9}(b), i.e., it always tends to increase the band gap. We are able to fit the band gap ($E_g$) of gated multilayer phosphorene as a function of the gate charge density ($n_g$) by using a simple polynomial $E_g=an_g^2+bn_g+c$ with $a$, $b$ and $c$ the polynomial coefficients. As an example, these three parameters are fitted for the case of dual (top and bottom) gates and they are listed in Table. \ref{tab1}. As can be seen, for all the cases of $N=2, 3, 4, 6$ layers, the presence of charge screening affects $a$ and $b$ significantly but has no effect on $c$. The values of $a$, $b$ and $c$ depend sensitively on the number of layers $N$. This polynomial fitting can be useful for the estimation of the band gap of gated multilayer phosphorene under ambient gate charge density (or gate electric field).

We note that in Ref. \cite{JhunElectronicstructurecharged2017}, similar results were obtained for this band-gap tuning in multilayer phosphorene by using first-principles calculations. However, the obtained results were restricted to the cases of bilayer and trilayer phosphorene and the screening effect was induced differently, i.e., by considering a charged system (our screening effect is induced by the gate electric field).


\subsection{Comparison with Experiment}

In order to verify the capability and accuracy of the present self-consistent TB model, we compare our theoretical band gaps of gated multilayer phosphorene with recent experimental results \cite{DengEfficientelectricalcontrol2017}. We consider multilayer phosphorene in the presence of dual (top and bottom) gates as in Ref. \cite{DengEfficientelectricalcontrol2017}, and take from this experimental work all the necessary parameters for theoretical modeling and calculations, including the number of layers and the displacement field. The experiment considered the charge-neutrality condition \cite{DengEfficientelectricalcontrol2017}, i.e., there are no excess carriers in the multilayer structure and thus the displacement field (denoted by $D$) is approximately uniform across the entire structure. This condition corresponds to our theoretical model where the charge densities of top and bottom gates are opposite in sign but equal in magnitude. By using the relation between the gate charge density and the gate electric field $F_{t,b}=e|n_{t,b}|/(2\varepsilon_0\kappa)$, the displacement fields of top and bottom gates can be written as $D_{t,b}=\kappa F_{t,b}=e|n_{t, b}|/(2\varepsilon_0)$. At the charge-neutrality condition ($n_t+n_b=0$), $D=D_t=D_b$ where $D$ is the displacement field of multilayer phosphorene, and we calculate the displacement-field tuning of the band gap of multilayer phosphorene, as was done experimentally in Ref. \cite{DengEfficientelectricalcontrol2017}. The magnitude of this band-gap tuning is computed as $\Delta E_g = E_g(D\neq0)-E_g(D=0)$.

\begin{figure}[htb!]
\begin{center}
\includegraphics[width=0.45\textwidth]{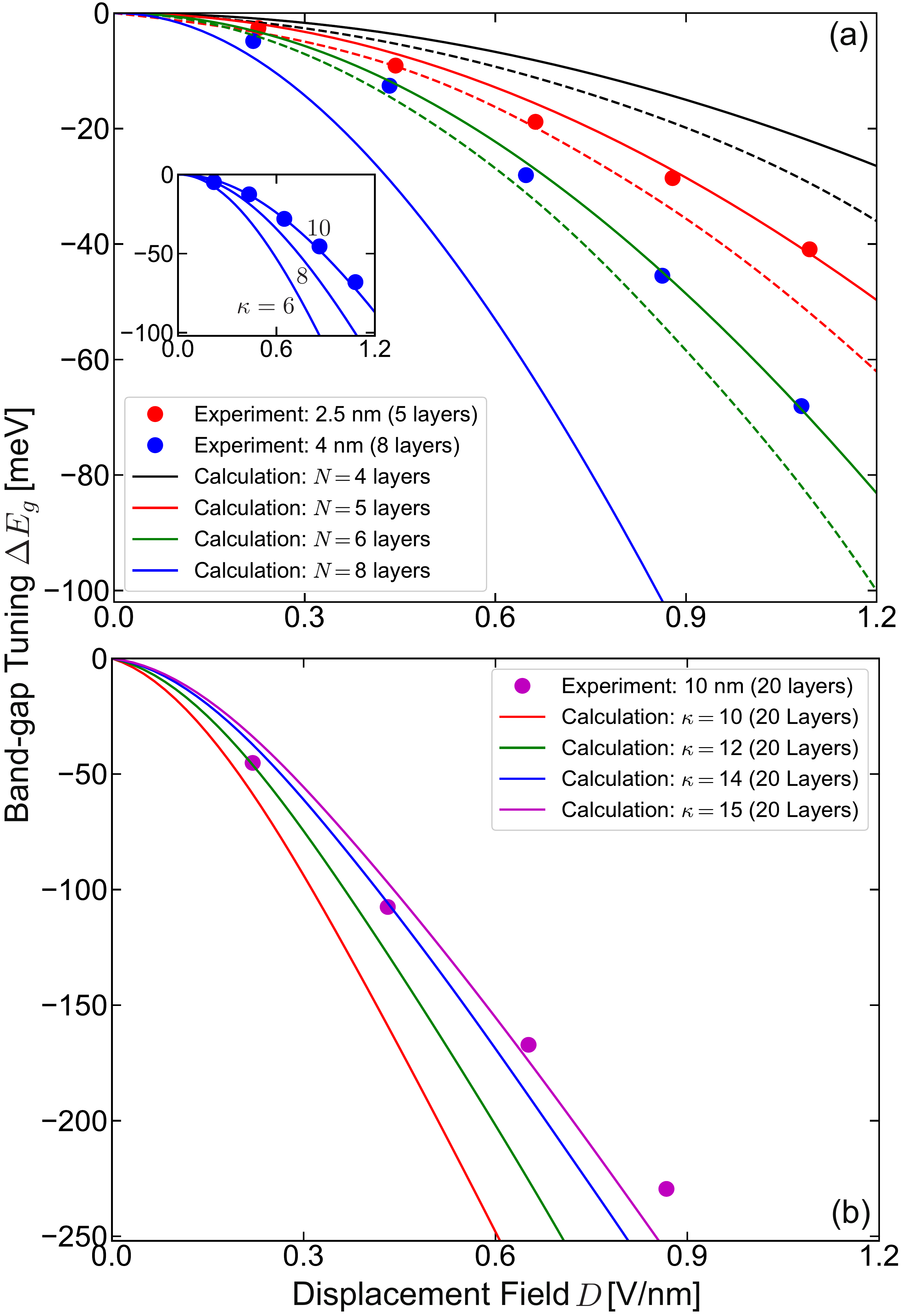}
\caption{Our theoretical results of the band-gap tuning $\Delta E_g$ of $N$-layer phosphorene ($N=4, 5, 6, 8, 20$) as a function of the displacement field $D$ (solid lines). Dashed lines and circle dots are, respectively, the theoretical and experimental results of Ref. \cite{DengEfficientelectricalcontrol2017} for $\Delta E_g$ as a function of $D$. The inset in (a) shows $\Delta E_g$ for the case of $N=8$ layers for different dielectric constants $\kappa$ as indicated.}\label{fig10}
\end{center}
\end{figure}

In Fig. \ref{fig10}(a), we show our theoretical results (solid lines) of $\Delta E_g$ as a function of $D$ for $N$-layer phosphorene ($N=4, 5, 6, 8$) and for comparative purposes, the theoretical (dashed lines) and experimental (circle dots) ones from Ref. \cite{DengEfficientelectricalcontrol2017}. As can be seen in Fig. \ref{fig10}(a), our theoretical results are in good agreement with the experimental ones for the 2.5 nm-thick sample (corresponding to $N=5$ layers), while the simple theoretical model presented in Ref. \cite{DengEfficientelectricalcontrol2017} overestimates the band-gap change with displacement field. However, the 4 nm-thick sample (corresponding to $N=8$ layers) exhibits a much smaller decrease in the band gap when compared with theory. The experimental results compare favorably with our results for $N=6$ layers. This discrepancy could possibly be due to the uncertainty in (i) the sample thickness in the experiment and (ii) the value of the dielectric constant of the sample. In spite of this discrepancy, our theoretical results are able to capture qualitatively the main feature of the band-gap variation with displacement field (i.e., the nonlinear decrease with increasing displacement field). Notice that by increasing the dielectric constant $\kappa$ from 6 to 10, we are able to fit the experimental results for the 4 nm-thick sample, as shown in the inset of Fig. \ref{fig10}(a).

The band-gap reduction with displacement field is even more pronounced for multilayer phosphorene with $N=20$ layers (corresponding to the 10-nm-thick sample), as shown in Fig. \ref{fig10}(b). In order to fit the experimental results, we have to further increase $\kappa$ from 10 to 15. This increase of the $\kappa$ value is reasonable because multilayer phosphorene with larger number of layers was shown to have a larger dielectric constant \cite{DengEfficientelectricalcontrol2017,
KumarThicknesselectricfielddependentpolarizability2016a}. The physical reason is that as the band gap decreases with increasing number of phosphorene layers, the multilayer phosphorene system becomes more metallic and so the dielectric screening of the system becomes stronger, which indicates an increased dielectric constant.
Therefore, with only one adjustable parameter (i.e., the dielectric constant $\kappa$) which is varied within reasonable values, our self-consistent TB calculations are able to reproduce the experimental results.

Moreover, our theoretical results are qualitatively consistent with the theoretical ones presented in Ref. \cite{DengEfficientelectricalcontrol2017}, as shown by the solid and dashed curves in Fig. \ref{fig10}(a). The quantitative difference between our model and the one used in Ref. \cite{DengEfficientelectricalcontrol2017} is due to the fact that: (i) the screening mechanism is different: ours is induced by the gate electric field while theirs was introduced by considering a band-gap-dependent dielectric constant; and (ii) our band gaps are obtained within a self-consistent Hartree scheme while their band gaps were obtained by self-consistently iterating a uniform dielectric constant of the multilayers via the band-gap dependence.

\section{Concluding Remarks}

We have theoretically investigated the electronic properties of multilayer phosphorene with top and/or bottom gates applied to produce a perpendicular electric field. By taking into account the electric-field-induced charge screening, a self-consistent tight-binding approach was employed to obtain the electronic band structure of gated multilayer phosphorene and the charge densities on the different phosphorene layers.

We found that in the presence of only a top (bottom) gate, the Fermi energy is located in the conduction (valence) band of multilayer phosphorene, which indicates a finite density of electrons (holes) in the system; whereas in the presence of both top and bottom
gates, the Fermi energy can be tuned into the band gap of multilayer phosphorene, which indicates no excess carriers in the system. The gate-induced carrier densities on the different phosphorene layers were found to increase with the gate charge density (or gate electric field) in either a linear or nonlinear fashion. We showed that the carrier densities on the different phosphorene layers are generally not equal to each other and thus produce an inhomogeneous (internal) electric field across the multilayer structure that counteracts the uniform (external) gate-produced one (i.e., the gate-induced charge screening).

Due to the puckered lattice structure, we found both intralayer and interlayer charge screening in gated multilayer phosphorene, which is different from gated multilayer graphene, where only interlayer charge screening is present. The gate-electric-field tuning of the band structure of multilayer phosphorene is distinctively different in the presence and absence of charge screening. For instance, we found that the unscreened band gap of multilayer phosphorene decreases dramatically with increasing electric-field strength. However, in the presence of charge screening, the magnitude of this band-gap decrease is significantly reduced and this reduction depends strongly on the number of phosphorene layers (which becomes more significant with increasing number of layers). Therefore, the critical strength of the gate electric field for the band-gap closure is significantly increased.

Moreover, we found charge density/screening anomalies in single-gated multilayer phosphorene due to the interlayer electronic coupling, and electron-hole bilayers in dual-gated multilayer phosphorene with tunable layer-dependent electron/hole densities.
Our theoretical results for the band-gap tuning (with gate electric field) agree with those obtained experimentally for multilayer phopshorene, thereby verifying the capability and accuracy of our self-consistent tight-binding approach.

\section{Acknowledgments}

This work was financially supported by the Flemish Science Foundation (FWO-Vl).

\bibliographystyle{apsrev4-1}

\begin{thebibliography}{31}%
\makeatletter
\providecommand \@ifxundefined [1]{%
 \@ifx{#1\undefined}
}%
\providecommand \@ifnum [1]{%
 \ifnum #1\expandafter \@firstoftwo
 \else \expandafter \@secondoftwo
 \fi
}%
\providecommand \@ifx [1]{%
 \ifx #1\expandafter \@firstoftwo
 \else \expandafter \@secondoftwo
 \fi
}%
\providecommand \natexlab [1]{#1}%
\providecommand \enquote  [1]{``#1''}%
\providecommand \bibnamefont  [1]{#1}%
\providecommand \bibfnamefont [1]{#1}%
\providecommand \citenamefont [1]{#1}%
\providecommand \href@noop [0]{\@secondoftwo}%
\providecommand \href [0]{\begingroup \@sanitize@url \@href}%
\providecommand \@href[1]{\@@startlink{#1}\@@href}%
\providecommand \@@href[1]{\endgroup#1\@@endlink}%
\providecommand \@sanitize@url [0]{\catcode `\\12\catcode `\$12\catcode
  `\&12\catcode `\#12\catcode `\^12\catcode `\_12\catcode `\%12\relax}%
\providecommand \@@startlink[1]{}%
\providecommand \@@endlink[0]{}%
\providecommand \url  [0]{\begingroup\@sanitize@url \@url }%
\providecommand \@url [1]{\endgroup\@href {#1}{\urlprefix }}%
\providecommand \urlprefix  [0]{URL }%
\providecommand \Eprint [0]{\href }%
\providecommand \doibase [0]{http://dx.doi.org/}%
\providecommand \selectlanguage [0]{\@gobble}%
\providecommand \bibinfo  [0]{\@secondoftwo}%
\providecommand \bibfield  [0]{\@secondoftwo}%
\providecommand \translation [1]{[#1]}%
\providecommand \BibitemOpen [0]{}%
\providecommand \bibitemStop [0]{}%
\providecommand \bibitemNoStop [0]{.\EOS\space}%
\providecommand \EOS [0]{\spacefactor3000\relax}%
\providecommand \BibitemShut  [1]{\csname bibitem#1\endcsname}%
\let\auto@bib@innerbib\@empty
\bibitem [{\citenamefont {Li}\ \emph {et~al.}(2014)\citenamefont {Li},
  \citenamefont {Yu}, \citenamefont {Ye}, \citenamefont {Ge}, \citenamefont
  {Ou}, \citenamefont {Wu}, \citenamefont {Feng}, \citenamefont {Chen},\ and\
  \citenamefont {Zhang}}]{LiBlackphosphorusfieldeffect2014}%
  \BibitemOpen
  \bibfield  {author} {\bibinfo {author} {\bibfnamefont {L.}~\bibnamefont
  {Li}}, \bibinfo {author} {\bibfnamefont {Y.}~\bibnamefont {Yu}}, \bibinfo
  {author} {\bibfnamefont {G.~J.}\ \bibnamefont {Ye}}, \bibinfo {author}
  {\bibfnamefont {Q.}~\bibnamefont {Ge}}, \bibinfo {author} {\bibfnamefont
  {X.}~\bibnamefont {Ou}}, \bibinfo {author} {\bibfnamefont {H.}~\bibnamefont
  {Wu}}, \bibinfo {author} {\bibfnamefont {D.}~\bibnamefont {Feng}}, \bibinfo
  {author} {\bibfnamefont {X.~H.}\ \bibnamefont {Chen}}, \ and\ \bibinfo
  {author} {\bibfnamefont {Y.}~\bibnamefont {Zhang}},\ }\href@noop {}
  {\bibfield  {journal} {\bibinfo  {journal} {Nat. Nanotechnol.}\ }\textbf
  {\bibinfo {volume} {9}},\ \bibinfo {pages} {372} (\bibinfo {year}
  {2014})}\BibitemShut {NoStop}%
\bibitem [{\citenamefont {Liu}\ \emph {et~al.}(2014)\citenamefont {Liu},
  \citenamefont {Neal}, \citenamefont {Zhu}, \citenamefont {Luo}, \citenamefont
  {Xu}, \citenamefont {Tom{\'a}nek},\ and\ \citenamefont
  {Ye}}]{LiuPhosphoreneUnexplored2D2014}%
  \BibitemOpen
  \bibfield  {author} {\bibinfo {author} {\bibfnamefont {H.}~\bibnamefont
  {Liu}}, \bibinfo {author} {\bibfnamefont {A.~T.}\ \bibnamefont {Neal}},
  \bibinfo {author} {\bibfnamefont {Z.}~\bibnamefont {Zhu}}, \bibinfo {author}
  {\bibfnamefont {Z.}~\bibnamefont {Luo}}, \bibinfo {author} {\bibfnamefont
  {X.}~\bibnamefont {Xu}}, \bibinfo {author} {\bibfnamefont {D.}~\bibnamefont
  {Tom{\'a}nek}}, \ and\ \bibinfo {author} {\bibfnamefont {P.~D.}\ \bibnamefont
  {Ye}},\ }\href@noop {} {\bibfield  {journal} {\bibinfo  {journal} {ACS Nano}\
  }\textbf {\bibinfo {volume} {8}},\ \bibinfo {pages} {4033} (\bibinfo {year}
  {2014})}\BibitemShut {NoStop}%
\bibitem [{\citenamefont {Lu}\ \emph {et~al.}(2014)\citenamefont {Lu},
  \citenamefont {Nan}, \citenamefont {Hong}, \citenamefont {Chen},
  \citenamefont {Zhu}, \citenamefont {Liang}, \citenamefont {Ma}, \citenamefont
  {Ni}, \citenamefont {Jin},\ and\ \citenamefont
  {Zhang}}]{LuPlasmaassistedfabricationmonolayer2014}%
  \BibitemOpen
  \bibfield  {author} {\bibinfo {author} {\bibfnamefont {W.}~\bibnamefont
  {Lu}}, \bibinfo {author} {\bibfnamefont {H.}~\bibnamefont {Nan}}, \bibinfo
  {author} {\bibfnamefont {J.}~\bibnamefont {Hong}}, \bibinfo {author}
  {\bibfnamefont {Y.}~\bibnamefont {Chen}}, \bibinfo {author} {\bibfnamefont
  {C.}~\bibnamefont {Zhu}}, \bibinfo {author} {\bibfnamefont {Z.}~\bibnamefont
  {Liang}}, \bibinfo {author} {\bibfnamefont {X.}~\bibnamefont {Ma}}, \bibinfo
  {author} {\bibfnamefont {Z.}~\bibnamefont {Ni}}, \bibinfo {author}
  {\bibfnamefont {C.}~\bibnamefont {Jin}}, \ and\ \bibinfo {author}
  {\bibfnamefont {Z.}~\bibnamefont {Zhang}},\ }\href@noop {} {\bibfield
  {journal} {\bibinfo  {journal} {Nano Res.}\ }\textbf {\bibinfo {volume}
  {7}},\ \bibinfo {pages} {853} (\bibinfo {year} {2014})}\BibitemShut {NoStop}%
\bibitem [{\citenamefont {Das}\ \emph {et~al.}(2014)\citenamefont {Das},
  \citenamefont {Zhang}, \citenamefont {Demarteau}, \citenamefont {Hoffmann},
  \citenamefont {Dubey},\ and\ \citenamefont
  {Roelofs}}]{DasTunableTransportGap2014}%
  \BibitemOpen
  \bibfield  {author} {\bibinfo {author} {\bibfnamefont {S.}~\bibnamefont
  {Das}}, \bibinfo {author} {\bibfnamefont {W.}~\bibnamefont {Zhang}}, \bibinfo
  {author} {\bibfnamefont {M.}~\bibnamefont {Demarteau}}, \bibinfo {author}
  {\bibfnamefont {A.}~\bibnamefont {Hoffmann}}, \bibinfo {author}
  {\bibfnamefont {M.}~\bibnamefont {Dubey}}, \ and\ \bibinfo {author}
  {\bibfnamefont {A.}~\bibnamefont {Roelofs}},\ }\href@noop {} {\bibfield
  {journal} {\bibinfo  {journal} {Nano Lett.}\ }\textbf {\bibinfo {volume}
  {14}},\ \bibinfo {pages} {5733} (\bibinfo {year} {2014})}\BibitemShut
  {NoStop}%
\bibitem [{\citenamefont {Qiao}\ \emph {et~al.}(2014)\citenamefont {Qiao},
  \citenamefont {Kong}, \citenamefont {Hu}, \citenamefont {Yang},\ and\
  \citenamefont {Ji}}]{QiaoHighmobilitytransportanisotropy2014a}%
  \BibitemOpen
  \bibfield  {author} {\bibinfo {author} {\bibfnamefont {J.}~\bibnamefont
  {Qiao}}, \bibinfo {author} {\bibfnamefont {X.}~\bibnamefont {Kong}}, \bibinfo
  {author} {\bibfnamefont {Z.-X.}\ \bibnamefont {Hu}}, \bibinfo {author}
  {\bibfnamefont {F.}~\bibnamefont {Yang}}, \ and\ \bibinfo {author}
  {\bibfnamefont {W.}~\bibnamefont {Ji}},\ }\href@noop {} {\bibfield  {journal}
  {\bibinfo  {journal} {Nat. Commun.}\ }\textbf {\bibinfo {volume} {5}},\
  \bibinfo {pages} {4475} (\bibinfo {year} {2014})}\BibitemShut {NoStop}%
\bibitem [{\citenamefont {Rodin}\ \emph {et~al.}(2014)\citenamefont {Rodin},
  \citenamefont {Carvalho},\ and\ \citenamefont
  {Castro~Neto}}]{RodinStrainInducedGapModification2014a}%
  \BibitemOpen
  \bibfield  {author} {\bibinfo {author} {\bibfnamefont {A.~S.}\ \bibnamefont
  {Rodin}}, \bibinfo {author} {\bibfnamefont {A.}~\bibnamefont {Carvalho}}, \
  and\ \bibinfo {author} {\bibfnamefont {A.~H.}\ \bibnamefont {Castro~Neto}},\
  }\href@noop {} {\bibfield  {journal} {\bibinfo  {journal} {Phys. Rev. Lett.}\
  }\textbf {\bibinfo {volume} {112}},\ \bibinfo {pages} {176801} (\bibinfo
  {year} {2014})}\BibitemShut {NoStop}%
\bibitem [{\citenamefont {Yuan}\ \emph {et~al.}(2015)\citenamefont {Yuan},
  \citenamefont {Liu}, \citenamefont {Afshinmanesh}, \citenamefont {Li},
  \citenamefont {Xu}, \citenamefont {Sun}, \citenamefont {Lian}, \citenamefont
  {Curto}, \citenamefont {Ye}, \citenamefont {Hikita}, \citenamefont {Shen},
  \citenamefont {Zhang}, \citenamefont {Chen}, \citenamefont {Brongersma},
  \citenamefont {Hwang},\ and\ \citenamefont
  {Cui}}]{YuanPolarizationsensitivebroadbandphotodetector2015a}%
  \BibitemOpen
  \bibfield  {author} {\bibinfo {author} {\bibfnamefont {H.}~\bibnamefont
  {Yuan}}, \bibinfo {author} {\bibfnamefont {X.}~\bibnamefont {Liu}}, \bibinfo
  {author} {\bibfnamefont {F.}~\bibnamefont {Afshinmanesh}}, \bibinfo {author}
  {\bibfnamefont {W.}~\bibnamefont {Li}}, \bibinfo {author} {\bibfnamefont
  {G.}~\bibnamefont {Xu}}, \bibinfo {author} {\bibfnamefont {J.}~\bibnamefont
  {Sun}}, \bibinfo {author} {\bibfnamefont {B.}~\bibnamefont {Lian}}, \bibinfo
  {author} {\bibfnamefont {A.~G.}\ \bibnamefont {Curto}}, \bibinfo {author}
  {\bibfnamefont {G.}~\bibnamefont {Ye}}, \bibinfo {author} {\bibfnamefont
  {Y.}~\bibnamefont {Hikita}}, \bibinfo {author} {\bibfnamefont
  {Z.}~\bibnamefont {Shen}}, \bibinfo {author} {\bibfnamefont {S.-C.}\
  \bibnamefont {Zhang}}, \bibinfo {author} {\bibfnamefont {X.}~\bibnamefont
  {Chen}}, \bibinfo {author} {\bibfnamefont {M.}~\bibnamefont {Brongersma}},
  \bibinfo {author} {\bibfnamefont {H.~Y.}\ \bibnamefont {Hwang}}, \ and\
  \bibinfo {author} {\bibfnamefont {Y.}~\bibnamefont {Cui}},\ }\href@noop {}
  {\bibfield  {journal} {\bibinfo  {journal} {Nat. Nanotechnol.}\ }\textbf
  {\bibinfo {volume} {10}},\ \bibinfo {pages} {707} (\bibinfo {year}
  {2015})}\BibitemShut {NoStop}%
\bibitem [{\citenamefont {Youngblood}\ \emph {et~al.}(2015)\citenamefont
  {Youngblood}, \citenamefont {Chen}, \citenamefont {Koester},\ and\
  \citenamefont {Li}}]{YoungbloodWaveguideintegratedblackphosphorus2015a}%
  \BibitemOpen
  \bibfield  {author} {\bibinfo {author} {\bibfnamefont {N.}~\bibnamefont
  {Youngblood}}, \bibinfo {author} {\bibfnamefont {C.}~\bibnamefont {Chen}},
  \bibinfo {author} {\bibfnamefont {S.~J.}\ \bibnamefont {Koester}}, \ and\
  \bibinfo {author} {\bibfnamefont {M.}~\bibnamefont {Li}},\ }\href@noop {}
  {\bibfield  {journal} {\bibinfo  {journal} {Nat. Photonics}\ }\textbf
  {\bibinfo {volume} {9}},\ \bibinfo {pages} {247} (\bibinfo {year}
  {2015})}\BibitemShut {NoStop}%
\bibitem [{\citenamefont {Tran}\ \emph {et~al.}(2014)\citenamefont {Tran},
  \citenamefont {Soklaski}, \citenamefont {Liang},\ and\ \citenamefont
  {Yang}}]{TranLayercontrolledbandgap2014}%
  \BibitemOpen
  \bibfield  {author} {\bibinfo {author} {\bibfnamefont {V.}~\bibnamefont
  {Tran}}, \bibinfo {author} {\bibfnamefont {R.}~\bibnamefont {Soklaski}},
  \bibinfo {author} {\bibfnamefont {Y.}~\bibnamefont {Liang}}, \ and\ \bibinfo
  {author} {\bibfnamefont {L.}~\bibnamefont {Yang}},\ }\href@noop {} {\bibfield
   {journal} {\bibinfo  {journal} {Phys. Rev. B}\ }\textbf {\bibinfo {volume}
  {89}},\ \bibinfo {pages} {235319} (\bibinfo {year} {2014})}\BibitemShut
  {NoStop}%
\bibitem [{\citenamefont {Castellanos-Gomez}\ \emph {et~al.}(2014)\citenamefont
  {Castellanos-Gomez}, \citenamefont {Vicarelli}, \citenamefont {Prada},
  \citenamefont {Island}, \citenamefont {Narasimha-Acharya}, \citenamefont
  {Blanter}, \citenamefont {Groenendijk}, \citenamefont {Buscema},
  \citenamefont {Steele}, \citenamefont {Alvarez}, \citenamefont {Zandbergen},
  \citenamefont {Palacios},\ and\ \citenamefont {van~der
  Zant}}]{CastellanosGomezIsolationcharacterizationfewlayer2014}%
  \BibitemOpen
  \bibfield  {author} {\bibinfo {author} {\bibfnamefont {A.}~\bibnamefont
  {Castellanos-Gomez}}, \bibinfo {author} {\bibfnamefont {L.}~\bibnamefont
  {Vicarelli}}, \bibinfo {author} {\bibfnamefont {E.}~\bibnamefont {Prada}},
  \bibinfo {author} {\bibfnamefont {J.~O.}\ \bibnamefont {Island}}, \bibinfo
  {author} {\bibfnamefont {K.~L.}\ \bibnamefont {Narasimha-Acharya}}, \bibinfo
  {author} {\bibfnamefont {S.~I.}\ \bibnamefont {Blanter}}, \bibinfo {author}
  {\bibfnamefont {D.~J.}\ \bibnamefont {Groenendijk}}, \bibinfo {author}
  {\bibfnamefont {M.}~\bibnamefont {Buscema}}, \bibinfo {author} {\bibfnamefont
  {G.~A.}\ \bibnamefont {Steele}}, \bibinfo {author} {\bibfnamefont {J.~V.}\
  \bibnamefont {Alvarez}}, \bibinfo {author} {\bibfnamefont {H.~W.}\
  \bibnamefont {Zandbergen}}, \bibinfo {author} {\bibfnamefont {J.~J.}\
  \bibnamefont {Palacios}}, \ and\ \bibinfo {author} {\bibfnamefont {H.~S.~J.}\
  \bibnamefont {van~der Zant}},\ }\href@noop {} {\bibfield  {journal} {\bibinfo
   {journal} {2D Mater.}\ }\textbf {\bibinfo {volume} {1}},\ \bibinfo {pages}
  {025001} (\bibinfo {year} {2014})}\BibitemShut {NoStop}%
\bibitem [{\citenamefont {Rudenko}\ and\ \citenamefont
  {Katsnelson}(2014)}]{RudenkoQuasiparticlebandstructure2014}%
  \BibitemOpen
  \bibfield  {author} {\bibinfo {author} {\bibfnamefont {A.~N.}\ \bibnamefont
  {Rudenko}}\ and\ \bibinfo {author} {\bibfnamefont {M.~I.}\ \bibnamefont
  {Katsnelson}},\ }\href@noop {} {\bibfield  {journal} {\bibinfo  {journal}
  {Phys. Rev. B}\ }\textbf {\bibinfo {volume} {89}},\ \bibinfo {pages} {201408}
  (\bibinfo {year} {2014})}\BibitemShut {NoStop}%
\bibitem [{\citenamefont {Lei}\ \emph {et~al.}(2016)\citenamefont {Lei},
  \citenamefont {Wang}, \citenamefont {Huang}, \citenamefont {Sun},\ and\
  \citenamefont {Zhang}}]{LeiStackingFaultEnriching2016}%
  \BibitemOpen
  \bibfield  {author} {\bibinfo {author} {\bibfnamefont {S.}~\bibnamefont
  {Lei}}, \bibinfo {author} {\bibfnamefont {H.}~\bibnamefont {Wang}}, \bibinfo
  {author} {\bibfnamefont {L.}~\bibnamefont {Huang}}, \bibinfo {author}
  {\bibfnamefont {Y.-Y.}\ \bibnamefont {Sun}}, \ and\ \bibinfo {author}
  {\bibfnamefont {S.}~\bibnamefont {Zhang}},\ }\href@noop {} {\bibfield
  {journal} {\bibinfo  {journal} {Nano Lett.}\ }\textbf {\bibinfo {volume}
  {16}},\ \bibinfo {pages} {1317} (\bibinfo {year} {2016})}\BibitemShut
  {NoStop}%
\bibitem [{\citenamefont {{\c C}ak{\i}r}\ \emph {et~al.}(2015)\citenamefont
  {{\c C}ak{\i}r}, \citenamefont {Sevik},\ and\ \citenamefont
  {Peeters}}]{CakirSignificanteffectstacking2015}%
  \BibitemOpen
  \bibfield  {author} {\bibinfo {author} {\bibfnamefont {D.}~\bibnamefont {{\c
  C}ak{\i}r}}, \bibinfo {author} {\bibfnamefont {C.}~\bibnamefont {Sevik}}, \
  and\ \bibinfo {author} {\bibfnamefont {F.~M.}\ \bibnamefont {Peeters}},\
  }\href@noop {} {\bibfield  {journal} {\bibinfo  {journal} {Phys. Rev. B}\
  }\textbf {\bibinfo {volume} {92}},\ \bibinfo {pages} {165406} (\bibinfo
  {year} {2015})}\BibitemShut {NoStop}%
\bibitem [{\citenamefont {Wang}\ \emph {et~al.}(2015)\citenamefont {Wang},
  \citenamefont {Liu}, \citenamefont {Kawazoe},\ and\ \citenamefont
  {Geng}}]{WangRoleInterlayerCoupling2015}%
  \BibitemOpen
  \bibfield  {author} {\bibinfo {author} {\bibfnamefont {V.}~\bibnamefont
  {Wang}}, \bibinfo {author} {\bibfnamefont {Y.~C.}\ \bibnamefont {Liu}},
  \bibinfo {author} {\bibfnamefont {Y.}~\bibnamefont {Kawazoe}}, \ and\
  \bibinfo {author} {\bibfnamefont {W.~T.}\ \bibnamefont {Geng}},\ }\href@noop
  {} {\bibfield  {journal} {\bibinfo  {journal} {J. Phys. Chem. Lett.}\
  }\textbf {\bibinfo {volume} {6}},\ \bibinfo {pages} {4876} (\bibinfo {year}
  {2015})}\BibitemShut {NoStop}%
\bibitem [{\citenamefont {Dai}\ and\ \citenamefont
  {Zeng}(2014)}]{DaiBilayerPhosphoreneEffect2014}%
  \BibitemOpen
  \bibfield  {author} {\bibinfo {author} {\bibfnamefont {J.}~\bibnamefont
  {Dai}}\ and\ \bibinfo {author} {\bibfnamefont {X.~C.}\ \bibnamefont {Zeng}},\
  }\href@noop {} {\bibfield  {journal} {\bibinfo  {journal} {J. Phys. Chem.
  Lett.}\ }\textbf {\bibinfo {volume} {5}},\ \bibinfo {pages} {1289} (\bibinfo
  {year} {2014})}\BibitemShut {NoStop}%
\bibitem [{\citenamefont {Rudenko}\ \emph {et~al.}(2015)\citenamefont
  {Rudenko}, \citenamefont {Yuan},\ and\ \citenamefont
  {Katsnelson}}]{Rudenkorealisticdescriptionmultilayer2015}%
  \BibitemOpen
  \bibfield  {author} {\bibinfo {author} {\bibfnamefont {A.~N.}\ \bibnamefont
  {Rudenko}}, \bibinfo {author} {\bibfnamefont {S.}~\bibnamefont {Yuan}}, \
  and\ \bibinfo {author} {\bibfnamefont {M.~I.}\ \bibnamefont {Katsnelson}},\
  }\href@noop {} {\bibfield  {journal} {\bibinfo  {journal} {Phys. Rev. B}\
  }\textbf {\bibinfo {volume} {92}},\ \bibinfo {pages} {085419} (\bibinfo
  {year} {2015})}\BibitemShut {NoStop}%
\bibitem [{\citenamefont {Fei}\ and\ \citenamefont
  {Yang}(2014)}]{FeiStrainEngineeringAnisotropicElectrical2014}%
  \BibitemOpen
  \bibfield  {author} {\bibinfo {author} {\bibfnamefont {R.}~\bibnamefont
  {Fei}}\ and\ \bibinfo {author} {\bibfnamefont {L.}~\bibnamefont {Yang}},\
  }\href@noop {} {\bibfield  {journal} {\bibinfo  {journal} {Nano Lett.}\
  }\textbf {\bibinfo {volume} {14}},\ \bibinfo {pages} {2884} (\bibinfo {year}
  {2014})}\BibitemShut {NoStop}%
\bibitem [{\citenamefont {Liu}\ \emph {et~al.}(2015)\citenamefont {Liu},
  \citenamefont {Zhang}, \citenamefont {Abdalla}, \citenamefont {Fazzio},\ and\
  \citenamefont {Zunger}}]{LiuSwitchingNormalInsulator2015}%
  \BibitemOpen
  \bibfield  {author} {\bibinfo {author} {\bibfnamefont {Q.}~\bibnamefont
  {Liu}}, \bibinfo {author} {\bibfnamefont {X.}~\bibnamefont {Zhang}}, \bibinfo
  {author} {\bibfnamefont {L.~B.}\ \bibnamefont {Abdalla}}, \bibinfo {author}
  {\bibfnamefont {A.}~\bibnamefont {Fazzio}}, \ and\ \bibinfo {author}
  {\bibfnamefont {A.}~\bibnamefont {Zunger}},\ }\href@noop {} {\bibfield
  {journal} {\bibinfo  {journal} {Nano Lett.}\ }\textbf {\bibinfo {volume}
  {15}},\ \bibinfo {pages} {1222} (\bibinfo {year} {2015})}\BibitemShut
  {NoStop}%
\bibitem [{\citenamefont {Yuan}\ \emph {et~al.}(2016)\citenamefont {Yuan},
  \citenamefont {{van Veen}}, \citenamefont {Katsnelson},\ and\ \citenamefont
  {Rold{\'a}n}}]{YuanQuantumHalleffect2016}%
  \BibitemOpen
  \bibfield  {author} {\bibinfo {author} {\bibfnamefont {S.}~\bibnamefont
  {Yuan}}, \bibinfo {author} {\bibfnamefont {E.}~\bibnamefont {{van Veen}}},
  \bibinfo {author} {\bibfnamefont {M.~I.}\ \bibnamefont {Katsnelson}}, \ and\
  \bibinfo {author} {\bibfnamefont {R.}~\bibnamefont {Rold{\'a}n}},\
  }\href@noop {} {\bibfield  {journal} {\bibinfo  {journal} {Phys. Rev. B}\
  }\textbf {\bibinfo {volume} {93}},\ \bibinfo {pages} {245433} (\bibinfo
  {year} {2016})}\BibitemShut {NoStop}%
\bibitem [{\citenamefont {Pereira}\ and\ \citenamefont
  {Katsnelson}(2015)}]{PereiraLandaulevelssinglelayer2015}%
  \BibitemOpen
  \bibfield  {author} {\bibinfo {author} {\bibfnamefont {J.~M.}\ \bibnamefont
  {Pereira}}\ and\ \bibinfo {author} {\bibfnamefont {M.~I.}\ \bibnamefont
  {Katsnelson}},\ }\href@noop {} {\bibfield  {journal} {\bibinfo  {journal}
  {Phys. Rev. B}\ }\textbf {\bibinfo {volume} {92}},\ \bibinfo {pages} {075437}
  (\bibinfo {year} {2015})}\BibitemShut {NoStop}%
\bibitem [{\citenamefont {Wu}\ \emph {et~al.}(2017)\citenamefont {Wu},
  \citenamefont {Chen}, \citenamefont {Gumbs},\ and\ \citenamefont
  {Lin}}]{WuFieldinduceddiversequantizations2017}%
  \BibitemOpen
  \bibfield  {author} {\bibinfo {author} {\bibfnamefont {J.-Y.}\ \bibnamefont
  {Wu}}, \bibinfo {author} {\bibfnamefont {S.-C.}\ \bibnamefont {Chen}},
  \bibinfo {author} {\bibfnamefont {G.}~\bibnamefont {Gumbs}}, \ and\ \bibinfo
  {author} {\bibfnamefont {M.-F.}\ \bibnamefont {Lin}},\ }\href@noop {}
  {\bibfield  {journal} {\bibinfo  {journal} {Phys. Rev. B}\ }\textbf {\bibinfo
  {volume} {95}},\ \bibinfo {pages} {115411} (\bibinfo {year}
  {2017})}\BibitemShut {NoStop}%
\bibitem [{\citenamefont {McCann}(2006)}]{McCannAsymmetrygapelectronic2006}%
  \BibitemOpen
  \bibfield  {author} {\bibinfo {author} {\bibfnamefont {E.}~\bibnamefont
  {McCann}},\ }\href@noop {} {\bibfield  {journal} {\bibinfo  {journal} {Phys.
  Rev. B}\ }\textbf {\bibinfo {volume} {74}},\ \bibinfo {pages} {161403}
  (\bibinfo {year} {2006})}\BibitemShut {NoStop}%
\bibitem [{\citenamefont {Avetisyan}\ \emph
  {et~al.}(2009{\natexlab{a}})\citenamefont {Avetisyan}, \citenamefont
  {Partoens},\ and\ \citenamefont
  {Peeters}}]{AvetisyanElectricfieldtuning2009}%
  \BibitemOpen
  \bibfield  {author} {\bibinfo {author} {\bibfnamefont {A.~A.}\ \bibnamefont
  {Avetisyan}}, \bibinfo {author} {\bibfnamefont {B.}~\bibnamefont {Partoens}},
  \ and\ \bibinfo {author} {\bibfnamefont {F.~M.}\ \bibnamefont {Peeters}},\
  }\href@noop {} {\bibfield  {journal} {\bibinfo  {journal} {Phys. Rev. B}\
  }\textbf {\bibinfo {volume} {79}},\ \bibinfo {pages} {035421} (\bibinfo
  {year} {2009}{\natexlab{a}})}\BibitemShut {NoStop}%
\bibitem [{\citenamefont {Avetisyan}\ \emph
  {et~al.}(2009{\natexlab{b}})\citenamefont {Avetisyan}, \citenamefont
  {Partoens},\ and\ \citenamefont
  {Peeters}}]{AvetisyanElectricfieldcontrolband2009}%
  \BibitemOpen
  \bibfield  {author} {\bibinfo {author} {\bibfnamefont {A.~A.}\ \bibnamefont
  {Avetisyan}}, \bibinfo {author} {\bibfnamefont {B.}~\bibnamefont {Partoens}},
  \ and\ \bibinfo {author} {\bibfnamefont {F.~M.}\ \bibnamefont {Peeters}},\
  }\href@noop {} {\bibfield  {journal} {\bibinfo  {journal} {Phys. Rev. B}\
  }\textbf {\bibinfo {volume} {80}},\ \bibinfo {pages} {195401} (\bibinfo
  {year} {2009}{\natexlab{b}})}\BibitemShut {NoStop}%
\bibitem [{\citenamefont {Koshino}\ and\ \citenamefont
  {McCann}(2009)}]{KoshinoGateinducedinterlayerasymmetry2009}%
  \BibitemOpen
  \bibfield  {author} {\bibinfo {author} {\bibfnamefont {M.}~\bibnamefont
  {Koshino}}\ and\ \bibinfo {author} {\bibfnamefont {E.}~\bibnamefont
  {McCann}},\ }\href@noop {} {\bibfield  {journal} {\bibinfo  {journal} {Phys.
  Rev. B}\ }\textbf {\bibinfo {volume} {79}},\ \bibinfo {pages} {125443}
  (\bibinfo {year} {2009})}\BibitemShut {NoStop}%
\bibitem [{\citenamefont {Zhang}\ \emph {et~al.}(2010)\citenamefont {Zhang},
  \citenamefont {Sahu}, \citenamefont {Min},\ and\ \citenamefont
  {MacDonald}}]{ZhangBandstructureABC2010}%
  \BibitemOpen
  \bibfield  {author} {\bibinfo {author} {\bibfnamefont {F.}~\bibnamefont
  {Zhang}}, \bibinfo {author} {\bibfnamefont {B.}~\bibnamefont {Sahu}},
  \bibinfo {author} {\bibfnamefont {H.}~\bibnamefont {Min}}, \ and\ \bibinfo
  {author} {\bibfnamefont {A.~H.}\ \bibnamefont {MacDonald}},\ }\href@noop {}
  {\bibfield  {journal} {\bibinfo  {journal} {Phys. Rev. B}\ }\textbf {\bibinfo
  {volume} {82}},\ \bibinfo {pages} {035409} (\bibinfo {year}
  {2010})}\BibitemShut {NoStop}%
\bibitem [{\citenamefont {Jhun}\ and\ \citenamefont
  {Park}(2017)}]{JhunElectronicstructurecharged2017}%
  \BibitemOpen
  \bibfield  {author} {\bibinfo {author} {\bibfnamefont {B.}~\bibnamefont
  {Jhun}}\ and\ \bibinfo {author} {\bibfnamefont {C.-H.}\ \bibnamefont
  {Park}},\ }\href@noop {} {\bibfield  {journal} {\bibinfo  {journal} {Phys.
  Rev. B}\ }\textbf {\bibinfo {volume} {96}},\ \bibinfo {pages} {085412}
  (\bibinfo {year} {2017})}\BibitemShut {NoStop}%
\bibitem [{\citenamefont {Deng}\ \emph {et~al.}(2017)\citenamefont {Deng},
  \citenamefont {Tran}, \citenamefont {Xie}, \citenamefont {Jiang},
  \citenamefont {Li}, \citenamefont {Guo}, \citenamefont {Wang}, \citenamefont
  {Tian}, \citenamefont {Koester}, \citenamefont {Wang}, \citenamefont {Cha},
  \citenamefont {Xia}, \citenamefont {Yang},\ and\ \citenamefont
  {Xia}}]{DengEfficientelectricalcontrol2017}%
  \BibitemOpen
  \bibfield  {author} {\bibinfo {author} {\bibfnamefont {B.}~\bibnamefont
  {Deng}}, \bibinfo {author} {\bibfnamefont {V.}~\bibnamefont {Tran}}, \bibinfo
  {author} {\bibfnamefont {Y.}~\bibnamefont {Xie}}, \bibinfo {author}
  {\bibfnamefont {H.}~\bibnamefont {Jiang}}, \bibinfo {author} {\bibfnamefont
  {C.}~\bibnamefont {Li}}, \bibinfo {author} {\bibfnamefont {Q.}~\bibnamefont
  {Guo}}, \bibinfo {author} {\bibfnamefont {X.}~\bibnamefont {Wang}}, \bibinfo
  {author} {\bibfnamefont {H.}~\bibnamefont {Tian}}, \bibinfo {author}
  {\bibfnamefont {S.~J.}\ \bibnamefont {Koester}}, \bibinfo {author}
  {\bibfnamefont {H.}~\bibnamefont {Wang}}, \bibinfo {author} {\bibfnamefont
  {J.~J.}\ \bibnamefont {Cha}}, \bibinfo {author} {\bibfnamefont
  {Q.}~\bibnamefont {Xia}}, \bibinfo {author} {\bibfnamefont {L.}~\bibnamefont
  {Yang}}, \ and\ \bibinfo {author} {\bibfnamefont {F.}~\bibnamefont {Xia}},\
  }\href@noop {} {\bibfield  {journal} {\bibinfo  {journal} {Nat. Commun.}\
  }\textbf {\bibinfo {volume} {8}},\ \bibinfo {pages} {14474} (\bibinfo {year}
  {2017})}\BibitemShut {NoStop}%
\bibitem [{\citenamefont {Zarenia}\ \emph {et~al.}(2017)\citenamefont
  {Zarenia}, \citenamefont {Neilson},\ and\ \citenamefont
  {Peeters}}]{ZareniaInhomogeneousphasescoupled2017}%
  \BibitemOpen
  \bibfield  {author} {\bibinfo {author} {\bibfnamefont {M.}~\bibnamefont
  {Zarenia}}, \bibinfo {author} {\bibfnamefont {D.}~\bibnamefont {Neilson}}, \
  and\ \bibinfo {author} {\bibfnamefont {F.~M.}\ \bibnamefont {Peeters}},\
  }\href@noop {} {\bibfield  {journal} {\bibinfo  {journal} {Sci. Rep.}\
  }\textbf {\bibinfo {volume} {7}},\ \bibinfo {pages} {11510} (\bibinfo {year}
  {2017})}\BibitemShut {NoStop}%
\bibitem [{\citenamefont {Moldovan}\ and\ \citenamefont
  {Peeters}(2016)}]{MoldovanPybinding2016}%
  \BibitemOpen
  \bibfield  {author} {\bibinfo {author} {\bibfnamefont {D.}~\bibnamefont
  {Moldovan}}\ and\ \bibinfo {author} {\bibfnamefont {F.~M.}\ \bibnamefont
  {Peeters}},\ }\href@noop {} {\bibfield  {journal} {\bibinfo  {journal}
  {\textit{Pybinding: a Python package for tight-binding calculations},
  \textit{http: //dx.doi.org/10.5281/zenodo.56818}}\ } (\bibinfo {year}
  {2016})}\BibitemShut {NoStop}%
\bibitem [{\citenamefont {Kumar}\ \emph {et~al.}(2016)\citenamefont {Kumar},
  \citenamefont {Bhadoria}, \citenamefont {Kumar}, \citenamefont {Bhowmick},
  \citenamefont {Chauhan},\ and\ \citenamefont
  {Agarwal}}]{KumarThicknesselectricfielddependentpolarizability2016a}%
  \BibitemOpen
  \bibfield  {author} {\bibinfo {author} {\bibfnamefont {P.}~\bibnamefont
  {Kumar}}, \bibinfo {author} {\bibfnamefont {B.~S.}\ \bibnamefont {Bhadoria}},
  \bibinfo {author} {\bibfnamefont {S.}~\bibnamefont {Kumar}}, \bibinfo
  {author} {\bibfnamefont {S.}~\bibnamefont {Bhowmick}}, \bibinfo {author}
  {\bibfnamefont {Y.~S.}\ \bibnamefont {Chauhan}}, \ and\ \bibinfo {author}
  {\bibfnamefont {A.}~\bibnamefont {Agarwal}},\ }\href@noop {} {\bibfield
  {journal} {\bibinfo  {journal} {Phys. Rev. B}\ }\textbf {\bibinfo {volume}
  {93}},\ \bibinfo {pages} {195428} (\bibinfo {year} {2016})}\BibitemShut
  {NoStop}%
\end{thebibliography}

%

\end{document}